\documentclass[3p,10pt, sort&compress, colorlinks, linkcolor=blue, citecolor=blue, urlcolor=blue]{article}

\usepackage[utf8]{inputenc}
\usepackage[sort&compress,numbers]{natbib}

\usepackage[utf8]{inputenc}

\usepackage{amssymb}
\usepackage{amsmath,mathrsfs}
\usepackage{empheq}

\usepackage{booktabs}

\usepackage[sc]{mathpazo}

\usepackage{color,soul}

\usepackage{extarrows}

\usepackage{verbatim}

\usepackage{lineno}

\usepackage{siunitx}

\usepackage{stfloats}

\usepackage{bm}

\usepackage{graphicx}

\usepackage{caption}
\usepackage{subcaption}

\usepackage{multicol}

\usepackage{multirow}

\usepackage{makecell}

\usepackage{mhchem}

\usepackage[toc,page]{appendix}

\usepackage{float}

\usepackage{empheq}

\usepackage{enumerate}

\usepackage{threeparttable}

\usepackage{tabularx}

\usepackage{sectsty}
\usepackage[toc,page]{appendix}
\usepackage[colorlinks,linkcolor=blue,anchorcolor=green,citecolor=blue]{hyperref}

\newcommand{\figref}[1]{Fig.~\ref{#1}}
\newcommand{\subfigref}[2]{Fig.~\ref{#1}\ce{#2}}
\newcommand{\subsfigref}[3]{Figs.~\ref{#1}\ce{#2} and \ref{#1}\ce{#3}}
\newcommand{\subssfigref}[3]{Figs.~\ref{#1}\ce{#2}-\ref{#1}\ce{#3}}
\newcommand{\tabref}[1]{Table.~\ref{#1}}
\renewcommand{\eqref}[1]{Eq.~$($\ref{#1}$)$}

\newcommand{\etal}{{\textit{et al.}}}

\usepackage{cuted}
\usepackage{flushend}

\usepackage{authblk}

\begin{document}
\def\smco{\ce{SmCo}\text{-1:7}}

\newcommand*{\diff}{\mathop{}\!\mathrm{d}}
\newcommand*{\Diff}{\mathop{}\!\mathrm{D}}

\newcommand*{\pd}[2]{\mathop{}\!\frac{\partial #1}{\partial #2}}
\newcommand*{\varid}[2]{\mathop{}\!\frac{\delta #1}{\delta #2}}
\newcommand*{\fed}[1]{f_\mathrm{#1}}


\def\fedf{\mathscr{F}}


\def\Rsq{\mathrm{R}^2}
\def\mse{\mathrm{MSE}}

\newcommand*{\E}[1]{\mathop{}\!\times 10^{#1}}

\def\deg{^{\circ}}
\def\CI{\text{CI}_{95\%}}

\def\vf15{\varphi_{S}}
\def\vfz{\varphi_\mathrm{Z}}

\def\ws{w_\mathrm{S}}
\def\wz{w_\mathrm{Z}}

\def\Ku{K_\mathrm{u}}
\def\Ms{M_\mathrm{s}}
\def\mv{\mathbf{m}}
\def\pos{\mathbf{r}}
\def\mag{\mathbf{m}}
\def\H{\mathbf{H}}
\def\MH{M\text{-}H}

\def\Ae{A_\mathrm{e}}
\def\sigdw{\sigma_\mathrm{dw}}
\def\ldw{l_\mathrm{dw}}

\def\Msz{\Ms^\text{(Z)}}
\def\sigdwz{\sigdw^\text{(Z)}}
\def\ldwz{\ldw^\text{(Z)}}

\def\Msm{\Ms^\text{(M)}}
\def\sigdwm{\sigdw^\text{(M)}}
\def\ldwm{\ldw^\text{(M)}}

\def\Mss{\Ms^\text{(S)}}
\def\sigdws{\sigdw^\text{(S)}}
\def\ldws{\ldw^\text{(S)}}

\def\Mssm{\Ms^\text{(S-M)}}
\def\sigdwsm{\sigdw^\text{(S-M)}}
\def\ldwsm{\ldw^\text{(S-M)}}

\def\Mszm{\Ms^\text{(Z-M)}}
\def\sigdwzm{\sigdw^\text{(Z-M)}}
\def\ldwzm{\ldw^\text{(Z-M)}}

\def\eau{\mathbf{u}}

\def\Hc{H_\mathrm{c}}
\def\Hp{H_\mathrm{p}}
\def\Hn{H_\mathrm{n}}
\def\Hani{H_\mathrm{a}}

\def\Pp{\Phi_\mathrm{p}}

\def\Hv{\mathbf{H}}
\def\Hvext{\mathbf{H}_\mathrm{ext}}
\def\Bv{\mathbf{B}}
\def\Mv{\mathbf{M}}

\def\Hdm{\mathbf{H}_\mathrm{dm}}

\def\oriang{\vartheta}

\def\bHc{\bar{H}_\mathrm{c}}

\def\delFdelm{\frac{\delta \mathcal{F}}{\delta \mathbf{m}}}

\def\nm{\si{nm}}
\def\um{\si{\micro m}}

\def\vfs{\varphi_\mathrm{S}}
\def\vfz{\varphi_\mathrm{Z}}

\def\iNN{i\text{-NN}}
\def\fNN{f\text{-NN}}
\def\fmod{\mathcal{F}}
\def\imod{\mathcal{I}}
\def\Yv{\boldsymbol{Y}}
\def\Xv{\boldsymbol{X}}
\def\wv{\boldsymbol{\omega}}
\def\nvv{\boldsymbol{\nu}}
	

	\title{
		 Coercivity influence of nanostructure in SmCo-1:7 magnets: Machine learning of high-throughput micromagnetic data}

	\author[1,*]{Yangyiwei Yang}
	\author[1]{Patrick Kühn}
	\author[1,*]{Mozhdeh Fathidoost}
	\author[2]{Esmaeil Adabifiroozjaei}
	\author[3]{Ruiwen Xie}
	\author[1]{Eren Foya}
	\author[4]{Dominik Ohmer}
	\author[5]{Konstantin Skokov}
	\author[2]{Leopoldo Molina-Luna}
 	\author[5]{Oliver Gutfleisch}
 	\author[3]{Hongbin Zhang}
	\author[1,*]{Bai-Xiang Xu}
	
	\affil[1]{\small Mechanics of Functional Materials Division, Institute of Materials Science, Technische Universit\"at Darmstadt, Darmstadt 64287, Germany}
	
	\affil[2]{\small Advanced Electron Microscopy Division, Institute of Materials Science, Technische Universit\"at Darmstadt, Darmstadt 64287, Germany}
	
	\affil[3]{Theory of Magnetic Materials Division, Institute of Materials Science, Technische Universit\"at Darmstadt, Darmstadt 64287, Germany}
	
	\affil[4]{VACUUMSCHMELZE GmbH \& Co. KG, 63450 Hanau, Germany}
	
	\affil[5]{Functional Materials (FM), Materials Science Department, Technical University of Darmstadt, 64287 Darmstadt, Germany}
	
	\affil[*]{Corresponding authors: \url{xu@mfm.tu-darmstadt.de} (Bai-Xiang Xu), \url{yangyiwei.yang@mfm.tu-darmstadt.de} (Yangyiwei Yang), \url{mozhdeh.fathidoost@mfm.tu-darmstadt.de} (Mozhdeh Fathidoost)}
	\date{}
	\maketitle
	\renewcommand\Authands{ and }

	\begin{abstract}

As a pinning-controlled permanent magnet, tailoring the cellular nanostructure of samarium–cobalt-based 1:7-type ($\smco$) magnets remains crucial for improving magnetic performance. Jointing forward and inverse machine learning models with the high-throughput micromagnetic simulations (42,300 runs), we identify the nanostructural and magnetic features that are most effective for coercivity, combining both nucleation and pinning mechanisms. Sensitivity analyses reveal that the 1:5-phase enhances coercivity by providing high anisotropy, and the Z-phase strengthens pinning through fluctuations in domain wall energy. Cu additions in the 1:5-phase significantly reduce coercivity, while Fe substitutions in the 2:17-phase modestly reduce coercivity but improve pinning locally and increase saturation magnetization. Among all examined features, magnetocrystalline misorientation emerges as the dominant factor. Finally, the framework enables the inverse design of nanostructures with prescribed coercivity, demonstrating a computationally cost-effective toolkit for guiding the performance tailoring of $\mathrm{SmCo}$ magnets.

	\end{abstract}

	
	
\section*{Introduction}
 
\begin{figure}[!t]
    \centering
    \includegraphics[width=17cm]{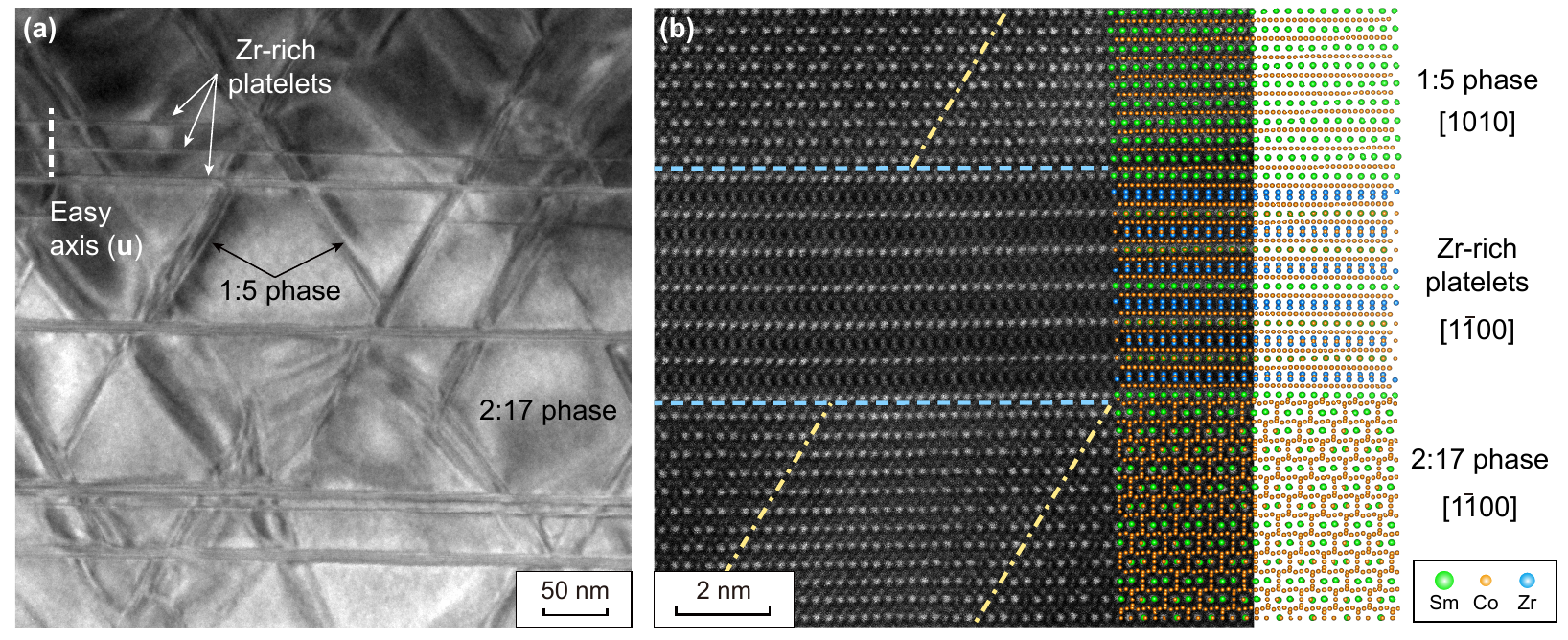}
    \caption{\small\textbf{The nanostructure of the SmCo-1:7 magnets}. (a) Bright-field TEM image of the $\smco$ magnets, where the nanostructure containing \ce{Sm(Co,Cu)5}, \ce{Sm2(Co,Fe)17} phases, and \ce{Zr}-rich platelet are presented. (b) HAADF-STEM images of a local junction among phases, where the atomic stacking matches correspondingly the theoretical superlattice of each phase.}
  \label{fig1}
\end{figure}

Samarium-cobalt-based 1:7-type magnet, usually formulated as \ce{Sm(Co,Fe,Cu,Zr)$_{7\pm\delta}$}, is known as a pinning-controlled permanent magnet that is capable of operating at elevated temperatures. Its corrosion endurance makes it indispensable for high-speed, high-power electric vehicles and aeronautic applications \cite{gutfleisch2011, strnat1991rare, liu1999microstructure, gutfleisch2006evolution}. Improving the operational performance of \ce{SmCo}-based 1:7-type magnet (hereinafter as $\smco$ magnet) receives continuous attention in the field of rare-earth-containing permanent magnets, because the potential of these materials, as characterized by crystalline anisotropy, is not fully utilized (known as the Brown paradox). Materials optimization relies on understanding the relationships between synthesis, microstructure, and performance. A sintered $\smco$ magnet presents a sophisticated microstructure with features spanning multiple length scales, ranging from the mesoscopic (1$\sim$100 \si{\micro m}) polycrystalline texture to the nanoscopic (1$\sim$100 \si{nm}) three-phase composite \cite{katter1996new, sepehri2017correlation, giron2023towards}. As shown in the \figref{fig1}a, this nanostructure takes a bi-pyramidal or diamond shape, consisting of a stripe-shaped \ce{Sm(Co,Cu)5} (hereinafter as 1:5-phase) and a cellular volume filled with \ce{Sm2(Co,Fe)17} (2:17-phase). Further subdivisions are created by the presence of Zr-rich platelet (Z-phase) developed perpendicular to the \textit{c}-axis \cite{song_cell-boundary-structure_2020}, intersecting both 2:17-phase and 1:5-phase in the diamond formation. By merit of high-resolution characterization like high-angle annular dark field scanning TEM (HAADF-STEM) imaging, one can observe the atomic stacking and coherent interfaces among these nanoscopic phases, as shown in \figref{fig1}b. As its length scale approaches the characteristic length of the magnetic domain-wall dynamics (notably the minimum domain-wall thickness among the constituent phases) and existing differences in domain-wall energy among conjugated phases, intersections of these nanoscopic phases are suggested as effective pinning sites, where the additional energy fluctuation is accompanied by further domain-wall motion \cite{tuprints20487}. Therefore, control and tailoring of such nanostructures form a key path leading to the engineering of ideal pinning sites. 

Optimizing the micro- and nanostructure to enhance the coercivity of $\smco$ permanent magnets is desired, particularly the geometric and magnetic properties of the nanoscopic phases, as it connects process parameters/conditions to the resulting local magnetic hysteresis. This delicate nanostructure is known as a collective outcome of its quinary chemical composition \cite{katter1996new, sepehri2017correlation, giron2023towards} and the thermal treatment \cite{kronmuller2003analysis, song_cell-boundary-structure_2020, matthias_tem-analysis_2002, pierobon2020, SONG2020, ZHOU20211560}. Among the alloying elements, Zr is crucial for forming the platelet-shaped Z-phase and stabilizing the cellular 2:17-phase \cite{tang2000effect, Hadjipanayis2000}, while Fe improves saturation magnetization and Cu provides concentration gradients in 1:5-phase \cite{yan2003microstructure, tuprints20487}. Variations in the chemical composition lead to changes not only in the geometry and topology of nanostructures but also in the magnetic properties of these nanoscopic phases \cite{lectard1994saturation}, ultimately affecting the strength of pinning sites. 

Although the overall coercivity of $\mathrm{SmCo}$ magnets is influenced by structural features across multiple length scales, particularly mesoscopic features such as grain boundaries and precipitates that locally reduce coercivity and often act as nuclei for reversed magnetic domains, the critical importance of engineering the nanoscopic cellular structure in designing $\mathrm{SmCo}$ magnets with optimal magnetic properties should not be overlooked. Specifically, the nanoscopic cellular structure directly establishes the energy barrier for domain-wall motion from multiple aspects, including, but not limited to, local magnetocrystalline anisotropy, concentration gradient, and short-range diffusion mechanism, distinguishing it from other mesoscopic features \cite{liu2023a, wang2022e, polin2025a}. Thus, the optimization of coercivity through the manipulation of nanostructures remains a highly active topic in the field \cite{liLamellarZrrichPhase2024, duerrschnabel2017, wang2022e, giron2023towards, polin2025a, hu2024a, liu2023a}. Notably, 
Polin \etal \cite{polin2025a} demonstrate that the cellular/lamellar nanostructure is critical for achieving high coercivity in $\smco$ magnets, as it provides the necessary nanoscale chemical profiles and pinning sites. While the Z-phase promotes the formation of this structure through twin-assisted nucleation, insufficient chemical partitioning (e.g., without Cu) results in poorly developed cells, resulting in low coercivity. Wang \etal \cite{wang2022e} further examined the coercivity effects of atomic-scale structural variants of the Z phase, namely hexagonal \ce{SmZr4Co13} and \ce{SmZr2Co9}, which are stabilized by Zr substitution and atomic displacements relative to the conventionally assumed rhombohedral \ce{(Sm,Zr)Co3} structure. Micromagnetic simulations reveal that the magnetocrystalline anisotropy of the Z phase governs the nucleation of magnetization reversal in the 2:17-phase, highlighting its critical role in coercivity and suggesting a pathway for performance enhancement through anisotropy tuning.

Many experimental works \cite{wang2018microstructure, maury1993, pierobon2020, YAN2019165459, tuprints20487, KRONMULLER2002545, GUTFLEISCH2006997, Liu1999MicrostructureAH, Zhang2001951223, Zhang2018TheEO, MATTHIAS20021353, YAN20041169, SONG2021, SONG2020, Tang1999, Jia2020DefectsaggregatedCB, Okabe2006, WANG2020156589, ZHOU20211560, Horiuchi2015} have defined characteristic quantities to describe phase formation and magnetic properties (notably, domain-wall thickness and energy, and saturation magnetization) for marking various samples with corresponding measured coercivities. Based on these works, two problems are posed: the forward problem concerns the identification of structural/magnetic features that are effective in achieving the desired coercivity; in the inverse problem, one aims to identify a specific nanostructure and end-member phases for a prescribed coercivity. In this context, the systematic knowledge extracted from an extensive dataset is of paramount importance in addressing these problems, aligning with the data-driven methodologies \cite{himanen2020}. To overcome the inefficient experimental-dominant trial-and-error methodologies, high-throughput micromagnetic simulations are employed to generate comprehensive and voluminous datasets. The solid physical foundation of micromagnetic simulations also aligns with the requirements of data-driven forward analyses and inverse design powered by machine learning (ML) techniques.  

In this work, we propose an ML-integrated data-driven framework for the systematic investigation and inverse design of nanostructured $\smco$ magnets. By combining high-throughput micromagnetic simulations with data-driven machine learning, we demonstrate how nanoscopic phases, through their geometry and intrinsic magnetic properties, influence the coercivity of the $\smco$ magnets through nucleation and domain-wall pinning mechanisms. Meanwhile, a strategy for the inverse design of nanostructures is deployed to efficiently predict corresponding nanostructures of a queried coercivity with pinning contribution. This framework not only identifies the most influential structural and micromagnetic descriptors, but also serves as a predictive tool for guiding experimental tailoring of $\smco$ nanostructures.

We begin with introducing the nanostructural and micromagnetic descriptors and their parameterization, informed by literature and experimental data. This is followed by forward machine learning of coercivity and its pinning contribution (quantified via a defined inhomogeneity factor) on more than 40,000 data points from high-throughput micromagnetic simulations. The inverse design of nanostructures from queried coercivity and inhomogeneity is further demonstrated. Building on this, we analyze how magnetocrystalline misorientation, along with the geometries and magnetic properties of nanoscopic phases, influences coercivity and the underlying pinning effect. Finally, sensitivity analyses are performed to assess and rank the effectiveness of these factors, including their contrasts among different phases.

\section*{Results}

\subsection*{Machine learning (ML) integrated data-driven framework}

\begin{figure}[!t]
    \centering
    \includegraphics[width=18cm]{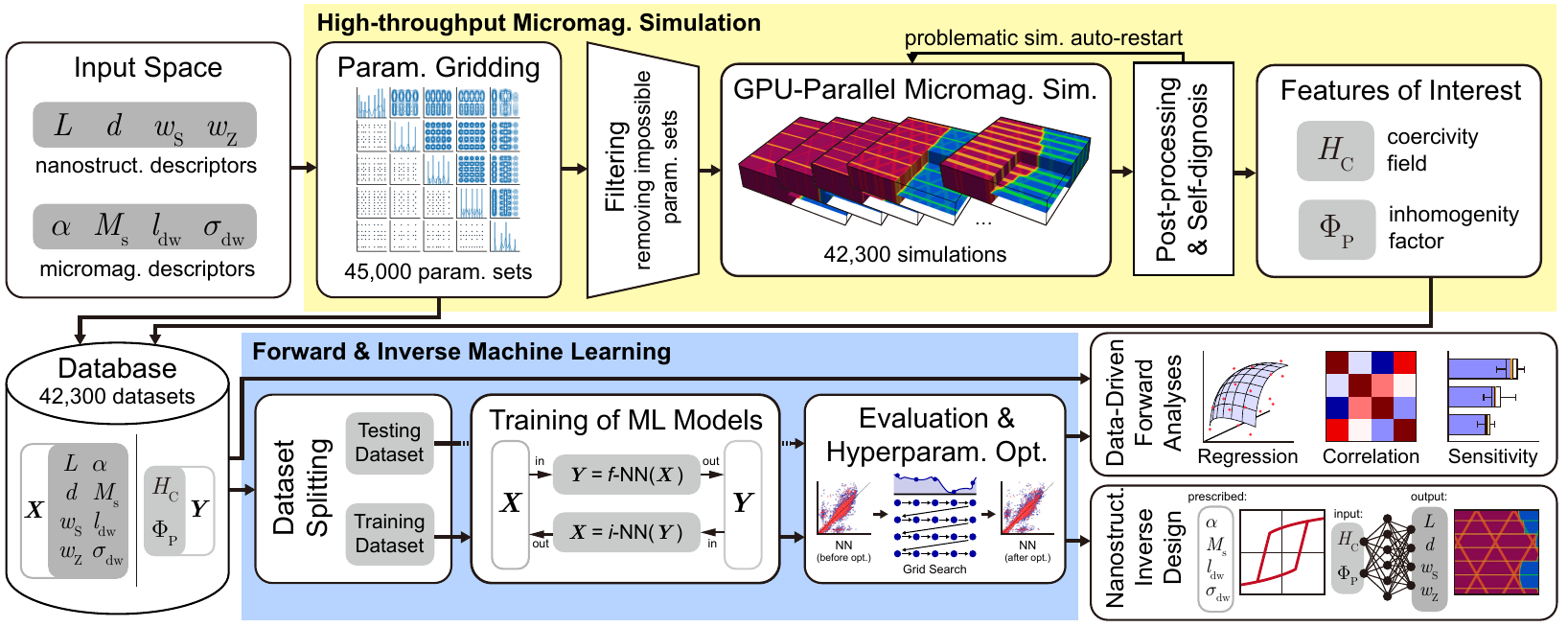}
    \caption{\small\textbf{Data-driven framework}. The proposed workflow integrates high-throughput micromagnetic simulations with forward and inverse machine learning to analyze the impact of nanostructure and phase properties on the coercivity of SmCo-1:7 permanent magnets.}
  \label{fig2}
\end{figure}

In \figref{fig2}, we present an ML-integrated, data-driven framework for investigating how the nanostructure and magnetic properties of underlying nanoscopic phases influence the coercivity of $\smco$ permanent magnets. This workflow incorporates parameterization and parameter gridding/filtering, GPU-parallelized micromagnetic simulations with automated result checking and restarting, and post-processing to extract features of the simulated hysteresis, such as the coercivity field $\Hc$ and the inhomogeneity factor $\Pp$ that characterizes the pinning contribution to the coercivity. The results of each simulation, along with the sampled parameters, are stored in a database for further use in training, evaluation, and hyperparameter optimization of ML models. These models facilitate data-driven forward analyses (e.g., regression, correlation, and sensitivity) and inverse design, enabling the specification of target magnetic properties and the prediction of corresponding optimal nanostructural descriptors. Details of these implemented methods are explained in the \textit{Methods} section. 

\subsection*{Input space: nanostructural and micromagnetic descriptors}

\begin{figure}[!t]
    \centering
    \includegraphics[width=18cm]{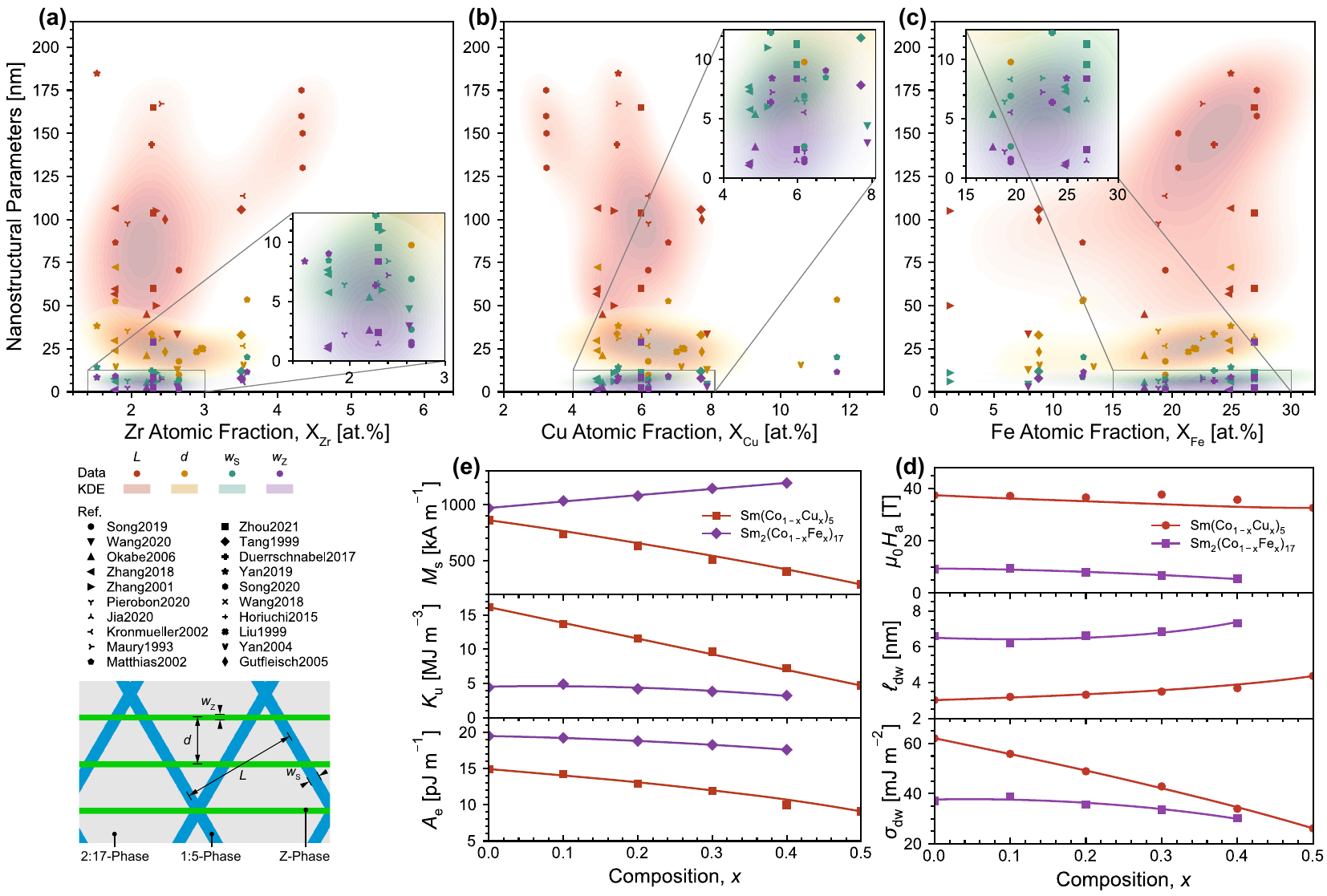}
    \caption{\small\textbf{Nanostructural descriptors and phase properties}. Distribution of the nanostructural descriptors vs. atomic fracture of (a) \ce{Zr} ($X_\mathrm{Zr}$), (b) \ce{Cu} ($X_\mathrm{Cu}$), and (c) \ce{Fe} ($X_\mathrm{Fe}$), from reviewed literatures \cite{wang2018microstructure, maury1993, pierobon2020, YAN2019165459, tuprints20487, KRONMULLER2002545, GUTFLEISCH2006997, Liu1999MicrostructureAH, Zhang2001951223, Zhang2018TheEO, MATTHIAS20021353, YAN20041169, SONG2021, SONG2020, Tang1999, Jia2020DefectsaggregatedCB, Okabe2006, WANG2020156589, ZHOU20211560, Horiuchi2015, duerrschnabel2017}. The composition dependency of magnetic properties of 1:5- and 2:17-phases are also presented: (d) $\Ae$, $\Ku$, $\Ms$, and (e) $\Hani$, $\ldw$, $\sigdw$. The experimental data is provided by \cite{liu2023a}.}
  \label{fig3}
\end{figure}

Following Katter \etal  \cite{katter1991new}, four descriptors are employed to characterize and recreate the nanostructural geometry: the average thickness of 1:5- ($w_\mathrm{S}$), the average thickness of Z-phase ($w_\mathrm{Z}$), the average interval of 1:5-phase ($L$) as well as that of Z-phases ($d$), as shown in the inset of \figref{fig2}. These four parameters were summarized from the characterization done in the existing experimental works \cite{wang2018microstructure, maury1993, pierobon2020, YAN2019165459, tuprints20487, KRONMULLER2002545, GUTFLEISCH2006997, Liu1999MicrostructureAH, Zhang2001951223, Zhang2018TheEO, MATTHIAS20021353, YAN20041169, SONG2021, SONG2020, Tang1999, Jia2020DefectsaggregatedCB, Okabe2006, WANG2020156589, ZHOU20211560, Horiuchi2015} (in Supplementary Table \hl{1}). Their distribution versus the atomic fraction Zr, Cu, and Fe is shown in \subssfigref{fig3}{a}{c}. The accessible compositional windows differ across the three alloying elements. Based on the 95\% KDE span, Zr shows the narrowest window ($X_\mathrm{Zr}^\mathrm{95\%}\sim[1.77$–$3.53]$ at.\%), Cu a slightly broader one ($X_\mathrm{Cu}^\mathrm{95\%}\sim[4.68$–$7.88]$ at.\%), while Fe spans the widest range ($X_\mathrm{Fe}^\mathrm{95\%}\sim[1.30$–$26.90]$ at.\%). For the 50\% KDE span, Zr and Cu windows change little ($X_\mathrm{Zr}^\mathrm{50\%}\sim[1.77$–$2.70]$, $X_\mathrm{Cu}^\mathrm{50\%}\sim[4.68$–$6.46]$ at.\%), but Fe narrows to $X_\mathrm{Fe}^\mathrm{50\%}\sim[15.90$–$26.90]$, consistent with most reported samples (\subfigref{fig3}{c}). Despite these differences, the four nanostructural descriptors converge to overlapping 95\% KDE ranges: $L^\mathrm{95\%}\sim[33.60,184.78]$, $d^\mathrm{95\%}\sim[9.78,72.28]$, $\ws^\mathrm{95\%}\sim[2.60,20.12]$, and $\wz^\mathrm{95\%}\sim[1.07–28.80]$ nm. Based on these, the sampling range of the nanostructure descriptors is $L\in[30,150]$, $d\in[10,55]$, $\ws\in[3,15]$, and $\wz\in[2, 10]$ nm, respectively. The pairplot of gridded nanostructural descriptors is presented in Supplementary Fig. \hl{1}.

Besides the nanostructural descriptors, we considered the influences of the micromagnetic descriptors of these nanoscopic phases on the resulting hysteresis of the nanostructure. These micromagnetic descriptors include the misorientation angle $\alpha$ (i.e., the angle between the applied magnetic field and the magnetocrystalline easy-axis), the saturation magnetization $\Ms$, the Bloch domain-wall thickness $\ldw$, and energy $\sigdw$. Here $\alpha$ varies from $0\deg$ (where the applied magnetic field $\Hvext$ field is along the easy-axis) to $90\deg$ (where $\Hvext$ is perpendicular to the easy-axis). It is worth noting that the exchange stiffness $\Ae$ and the magnetocrystalline anisotropy constant $\Ku$ for the uniaxial symmetry can be related to $\ldw$ and $\sigdw$ by 
 \begin{equation}
\sigdw=4\sqrt{\Ae\Ku}\quad\mathrm{and}\quad\ldw=\pi\sqrt{\frac{\Ae}{\Ku}}
 \end{equation}
with SI units. These intrinsic magnetic properties are influenced by the contents of alloying elements, as overviewed in the \textit{Introduction}. In \subfigref{fig3}{d} we present the composition dependency of $\Ae$, $\Ku$, and $\Ms$ of 1:5- and 2:17-phases based on the experimental measurements by Liu \etal  \cite{liu2023a} at room temperature. For $\Ms$, the 2:17-phase shows higher values that rise slightly with Fe content ($\sim 1000$–$1200$ \si{kA m^{-1}}), while the 1:5-phase is lower and decreases strongly with Cu content ($\sim 900$–$300$ \si{kA m^{-1}}). For $\Ku$, the 1:5-phase exhibits larger values that drop with Cu addition ($\sim 16$–$5$ \si{MJ m^{-3}}), peaking at \ce{Sm(Co_{0.9}Cu_{0.1})}, whereas the 2:17-phase stays nearly constant at a lower level ($\sim 3$–$5$ \si{MJ m^{-3}}). A similar trend appears for $\Ae$: the 1:5-phase decreases from $\sim 15$ to 9 \si{pJ m^{-1}}, while the 2:17-phase remains slightly higher but relatively stable ($\sim 19$–$17$ \si{pJ m^{-1}}).

\subfigref{fig3}{d} highlights that the 1:5-phase provides high magnetocrystalline anisotropy rather than magnetization, while the 2:17-phase contributes higher magnetization and exchange stiffness with modest magnetocrystalline anisotropy. Nonetheless, compared to $\Ae$ and $\Ku$, $\sigdw$ and $\ldw$ are the direct features of the domain wall and can reflect the pinning strength of the corresponding phase. Meanwhile, to analyze the contribution of these nanoscopic phases to the intrinsic coercivity and the effects of the compositions, here we also calculate the magnetocrystalline anisotropic field as
\begin{equation}
    \Hani=\frac{2\Ku}{\mu_0\Ms},\label{eq:Ha}
\end{equation}
which is treated ideally as the upper-bound coercivity of a homogeneous bulk material as of the Stoner-Wohlfarth model without contribution from the shape anisotropy \cite{stoner1948mechanism, coey2010magnetism}. Thus, the composition dependency of these dependent magnetic properties is calculated and presented in \subfigref{fig3}{e}. The 1:5-phase exhibits consistently higher $\Hani$ in the examined composition region with only a slight decrease with the Cu content ($\sim38–32$ \si{T}), whereas the 2:17-phase remains comparably lower ($\sim9–5$ \si{T}). On the other hand, the 1:5-phase maintains relatively narrow walls ($\sim3$ to 6 \si{nm}), while the 2:17-phase has broader walls that increase slightly with Fe content ($\sim6$ to 7 \si{nm}). Although the domain-wall thickness seems comparably less composition sensitive, the domain-wall energy $\sigdw$ of the 1:5-phase decreases drastically with Cu content from a relatively high value ($\sim62$ \si{mJ~m^{-2}}), while the 2:17-phase's remains at relatively lower values ($\sim39\text{ to }30$ \si{mJ~m^{-2}}) and reaches its maximum at \ce{Sm_2(Co_{0.9}Fe_{0.1})_17}. Within the composition range of $x\leq 0.4$, the 1:5-phase has a relatively higher $\sigdw$ compared to the 2:17-phase. This means that an extra energy gain is required for a domain wall to penetrate from the cellular 2:17-phase to the stripe-shaped 1:5-phase. In other words, the interface between the 2:17- and 1:5-phases can be considered a repulsive pinning site for magnets, a focus of research on $\smco$ magnets.

\begin{table}[ht]
\centering\small
\caption{Micromagnetic parameters for the nanoscopic phases in this work.}
\begin{tabular}{lcccccccccccccc}\hline                         
\multirow{2}{*}{}  & \multirow{2}{*}{Unit}      & \multicolumn{4}{c}{2:17-Phase, \ce{Sm2(Co_{1-x}Fe_{x})_{17}}} & \multicolumn{4}{c}{1:5-Phase, \ce{Sm(Co_{1-x}Cu_{x})_5}} & \multicolumn{3}{c}{Zr-rich}  \\ 
\cmidrule(lr){3-6} \cmidrule(lr){7-10} \cmidrule(lr){11-13}& &
$0.0$ & $0.1$  & $0.2$ & $0.3$ &
$0.0$ & $0.1$ & $0.2$ & $0.3$ &
Lit. & DFT & Mod.\\
\hline
$\Ae$ & \si{pJ~m^{-1}}          & 
19.48 &  19.21  & 18.80 &  18.27 &
14.92  &  14.04  &  13.10  &  11.98  &   
0.48   &  4.83 & 0.96 \\
$\Ku$ & \si{MJ~m^{-3}}     &
4.55  &  4.60  &  4.39  &  3.92   & 
16.14 &  13.85 &  11.55 &  9.26   &  
2.10  &  2.10 & 1.04 \\
$\Ms$ & \si{kA~m^{-1}} & 
971.41   &  1029.21  &  1085.75  &  1141.03  &
862.21  &  765.41  &  658.02  &  544.90 & 
310.4   & 290.0 & 310.4 \\
$\ldw$ &\si{nm}  & 
6.50  &  6.42  &  6.50  &  6.78   &
3.02  &  3.16  &  3.35  &  3.57   & 
1.50  & 4.76 & 3.02\\
$\sigdw$ & \si{mJ~m^{-2}}  & 
37.65  &  37.60  &  36.34  & 33.84 &
62.09  &  55.77  &  49.20  & 42.14 &
4.00   & 12.74 & 4.00 \\
$\mu_0\Hani$ & T &
9.36  &  8.94  &  8.09  &  6.87  &
37.45 &  36.18 &  35.11 &  34.00 &
13.53 &  14.48 &  6.70 \\
\hline
\end{tabular}
\label{tab:1}
\end{table}

In this work, the compositions of Cu and Fe were gridded as $x \in [0, 0.1, 0.2, 0.3]$ in the 1:5- and 2:17-phases, corresponding to atomic fractions of $X^\mathrm{P}_\mathrm{Cu} \in [0.0, 8.3, 16.7, 25.0]\text{ at.\%}$ and $X^\mathrm{P}_\mathrm{Fe} \in [0.0, 9.0, 17.9, 26.8]\text{ at.\%}$ within the respective phases. Note that the superscript `P' denotes the atomic fractions obtained within the phases, which are distinguished from those of the entire sample (see \subssfigref{fig3}{a}{c}). The associated micromagnetic descriptors were calculated accordingly, as summarized in Table~\ref{tab:1}. These ranges were chosen based on the peak Cu concentration in the 1:5-phase ($\sim26\text{ at.\%}$) and the peak Fe concentration in the 2:17-phase ($\sim27\text{ at.\%}$) observed by APT measurements~\cite{giron2023towards}. Comparable concentration profiles of Cu and Fe in the respective phases have also been reported by Duerrschnabel \etal ~\cite{duerrschnabel2017}.  
As for the Z-phase, it is generally regarded as an intermetallic compound with a fixed stoichiometry \cite{zhang2023, wang2022e,  decampos2007}. Thus, the formation and thickness of the Z-phase are directly linked to the Zr content. However, uncertainties remain regarding the atomic structure of the Z-phase and its influence on magnetic properties. It is conventionally regarded as having a rhombohedral \ce{(Sm,Zr)Co_3} (1:3-R) structure\cite{decampos2007}. On the other hand, recent research has proposed hexagonal \ce{(SmZr_4)Co_13} or \ce{(SmZr_2)Co_9} structures, which are more thermodynamically stable and have a higher $\Ku$ \cite{zhang2023}. In this work, we regarded the Z-phase as a 1:3-R phase, more specifically $\ce{(SmZr_2)Co_9}$. This is supported by the comparison between the HAADF-STEM images (\figref{fig1}b) and the relaxed lattice structure by density functional theory (DFT) calculation (details are summarized in the \textit{Method} section). To include the uncertainty in magnetic properties, three types of micromagnetic descriptors of Z-phase are examined, including values from the literature \cite{katter1996new} (labelled as `Lit.'), values from DFT calculations with the composition $\ce{(SmZr_2)Co_9}$ (labelled as `DFT'), and values modified assuming an identical $\ldw$ as 1:5-phase while taking $\sigdw$ and $\Ms$ from Ref. \cite{katter1996new} (labelled as `Mod.'). This can shed insight into how sensitive the domain-wall motion is to the magnetic properties in three nanoscopic phases, whose boundaries with other phases are expected to form effective pinning sites in the magnet. 
 
Based on the defined ranges of nanostructural and micromagnetic descriptors, the input space was constructed through parameter gridding (45,000 sets) and subsequently refined by filtering out non-physical combinations, for instance $\ws>L$ and/or $\wz>d$. This yields a total of 42,300 valid input parameter sets for the subsequent high-throughput GPU-parallelized micromagnetic simulations.

\subsection*{High-throughput micromagnetic simulations of magnetization reversal}

\begin{figure}[!t]
	\centering
 	\includegraphics[width=18cm]{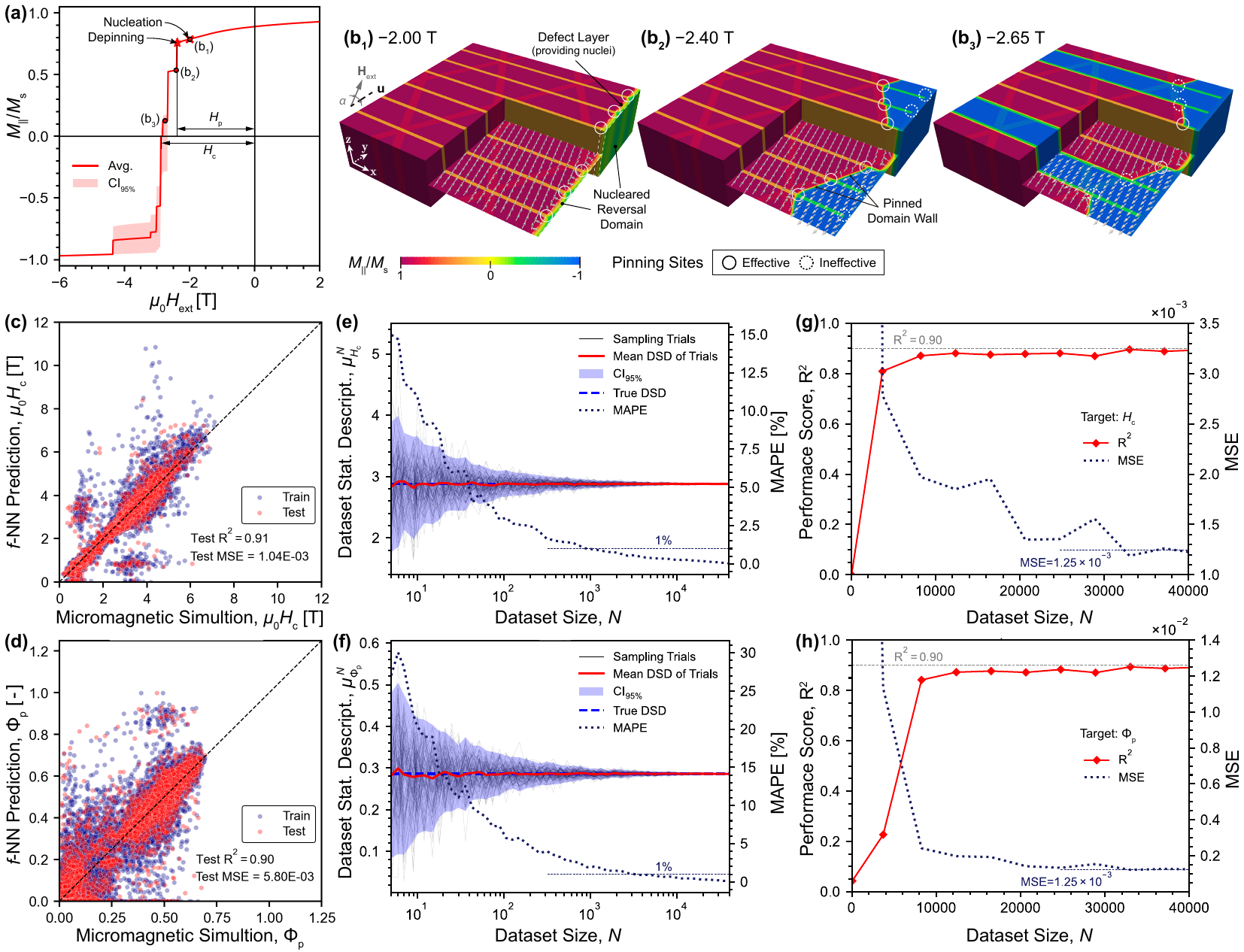}
	\caption{\small\textbf{Simulated hysteresis and forward neural networks
    }. (a) Average half-cycle hysteresis and 95\% confidence interval ($\CI$) by micromagnetic simulation with ten reversal processes. (b$_1$)-(b$_2$) Transient domain structures with marked pinning sites at selected applied field strength. Parity plot of the $\fNN$ prediction vs. micromagnetic simulation for (c) coercivity $\Hc$ and (d) inhomogeneity factor $\Pp$. Convergence analysis on the dataset statistical descriptors (DSD) for (e) $\Hc$ and (f) $\Pp$ vs. various dataset sizes $N$. The performance score of the machine learning for (g) $\Hc$ and (h) $\Pp$ vs various dataset sizes $N$. }
	\label{fig4}
\end{figure}

A finite-difference (FD) simulation domain containing three nanoscopic phases was constructed for each input set of nanostructural descriptors ($L$, $d$, $\ws$, $\wz$), with the corresponding micromagnetic descriptors ($\alpha$, $\Ae$, $\Ku$, $\Ms$) assigned to their respective spatial regions. To meet the requirements of the high-throughput scenario while reducing variations in the simulated hysteresis loops, a relatively efficient iterative minimization algorithm, such as the steepest conjugate gradient (SCG), was employed in the micromagnetic simulations, combined with an averaging procedure over multiple demagnetization cycles. Accordingly, each micromagnetic simulation comprised five to ten demagnetization half-cycles, in which the magnetization was reversed once from the positive saturation state to the negative state. 
Meanwhile, to emulate the effects of mesoscopic structures (such as polycrystalline texture, grain boundaries, and amorphous phases) and defects that can significantly reduce the coercivity by providing nucleation sites for magnetization reversal~\cite{staab2024,bance2014a,chen2018b,Yi2016a}, a defect layer was introduced at the boundary of the simulation domain, as shown in \subfigref{fig4}{b1}. This layer was modeled as a soft ferromagnetic phase with the same $\Ae$ and $\Ms$ as the matrix 2:17-phase but with a reduced anisotropy constant $\Ku' = 0.01\Ku$. 
Detailed theory and setup of high-throughput micromagnetic simulations are provided in the \textit{Method} section.

\figref{fig4}{a} presents the averaged half-cycle hysteresis curve with selected magnetization configurations at applied fields $\Hvext$, where the 95\% confidence interval ($\CI$) indicates numerical variations among simulated curves. The selected nanostructure has $L=100$, $d=40$, $\ws=10$, and $\wz= 5$ \si{nm}. The 1:5-phase takes $\ce{SmCo5}$, the 2:17-phase takes $\ce{Sm2Co17}$, and Z-phase uses Mod. parameters. The $\alpha=30\deg$ is set to demonstrate a reversal process with the existence of magnetocrystalline misorientation. The pinning–depinning events appear as step-like transitions of the hysteresis, with three representative domain structures shown in \subssfigref{fig4}{b1}{b3}. Following previous works~\cite{bance2014a}, the nucleation field $\Hn$ is defined by the formation of a minimum $90\deg$ domain wall at the defect layer (\subfigref{fig4}{b1}). The first discontinuity in the averaged curve is interpreted as the depinning field $\Hp$, corresponding to the minimum field required to overcome the energy barrier of the initial pinning sites. The migrated domain wall soon encounters pinning sites at the junctions among 2:17-, 1:5- and Z-phases, and becomes temporarily stabilized (\subfigref{fig4}{b2}). As the external field reaches $-2.65~\si{T}$, some of the pinning sites get penetrated again along with the rapid migration of the domain wall. To quantify the coercivity contribution from the pinning effects, a normalized inhomogeneity factor is defined
\begin{equation}
    \Pp = \frac{\Hc-\Hp}{\Hc}.\label{eq:Pp}
\end{equation}
In a homogeneous nanostructure, $\Hc$ generally equals $\Hp$, stating a non-impeded domain-wall motion after nucleation. When the nanostructure is inhomogeneous (like this $\smco$ nanostructure), $\Hc$ is further extended from $\Hp$ due to the impeded domain-wall motion by pinning sites.

\subsection*{Forward machine learning of coercivity with pinning contributions}

As the properties of interest (POI), $\Hc$ and the corresponding $\Pp$ were automatically determined from a total of 42,300 averaged $\MH$ curves resulting from micromagnetic simulations with an in-house multi-process workflow. This workflow also includes utilities to identify numerical errors and perform automatic recalculation, thereby preventing contamination of the dataset. After collecting the fully compliant simulation results and curating the dataset, a subset of the complete database was partitioned into training (80\%) and test (20\%) datasets for the development of a forward ML model, which provides the foundation for the subsequent data-driven analyses and inverse design. After evaluating six different ML models and algorithms (details in the \textit{Method} section), the neural networks (NN) with a multilayer perceptron architecture were selected for forward prediction (hereinafter denoted as $\fNN$), owing to their comparatively good performance and their ability to perform continuous inter- and extrapolations.

\subsfigref{fig4}{a}{b} present parity plots of $\fNN$ predictions versus\ micromagnetic simulations for POIs after hyperparameter optimization. For $\Hc$, most data points cluster around the 45° diagonal, indicating accurate predictions in the main data range. The distributions of training and test data points are similar, indicating no severe overfitting. With $\Rsq=0.91$ and $\text{MSE}=1.04\E{-3}$, the model achieves high explained variance and low error. Deviations occur in the high-value region ($\mu_0\Hc > 4$ T), where predictions are systematically underestimated. This is possibly due to the insufficient data in the high-value region. In the low-to-intermediate range ($\mu_0\Hc<4$ T), on the other hand, predictions align closely with ground truth. For $\Pp$, performance is comparable ($\Rsq=0.90$, $\text{MSE}=5.80\E{-3}$), though high-value cases ($\Pp>0.5$) again show underestimation and larger fluctuations. Overall, the $\fNN$ demonstrates strong predictive capability but exhibits a conservative bias in high-value regions.

To ensure that the dataset size $N$ is sufficient to reflect the characteristics of the complete dataset statistically, \subsfigref{fig4}{e}{f} illustrates the convergence of dataset statistical descriptors (DSDs) along with $N$ for $\Hc$ and $\Pp$, respectively. For each $N$, 100 independent sampling trials were performed within the input parameter spaces. The corresponding arithmetic means of resulting $\Hc$ and $\Pp$, denoted as $\mu_{\Hc}^N$ and $\mu_{\Pp}^N$, were examined and taken as DSDs (light black lines). For small $N$, large fluctuations and wide $\CI$ are observed, reflecting the instability of statistical estimates due to undersampled datasets. With increasing $N$, both the mean DSD of all trails (red line) and $\CI$ rapidly converge toward the true DSD of the complete dataset (in total 42,300 data points), which indicates that the sampled datasets increasingly capture the statistical characteristics of the full dataset. The mean absolute percentage error (MAPE) further demonstrates this convergence trend, approaching values below 1\% when $N\geq1,000$ for $\Hc$ and $N\geq3,000$ for $\Pp$. In this regard, the training dataset size of 33,840 is more than sufficient.

The performance of the trained $\fNN$ with respect to varying dataset size $N$ is shown in \subsfigref{fig4}{g}{h}. For both $\Hc$ and $\Pp$, the coefficient of determination $\Rsq$ increases rapidly with $N$ and stabilizes around 0.90 once $N\geq 10,000$. Meanwhile, the mean squared error (MSE) decreases sharply in the small-$N$ region and reaches around $1.25\E{-3}$ when $N \sim 40{,}000$. Beyond this point, further increasing the dataset size provides no significant improvement in either $\Rsq$ or MSE, indicating that the model performance saturates once sufficient data are available. These results confirm that the chosen training dataset size of 33,840 is well within the converged region, ensuring reliable predictive performance.

\subsection*{Inverse design of $\smco$ nanostructure}
   
\begin{figure}[!t]
	\centering
 	\includegraphics[width=18cm]{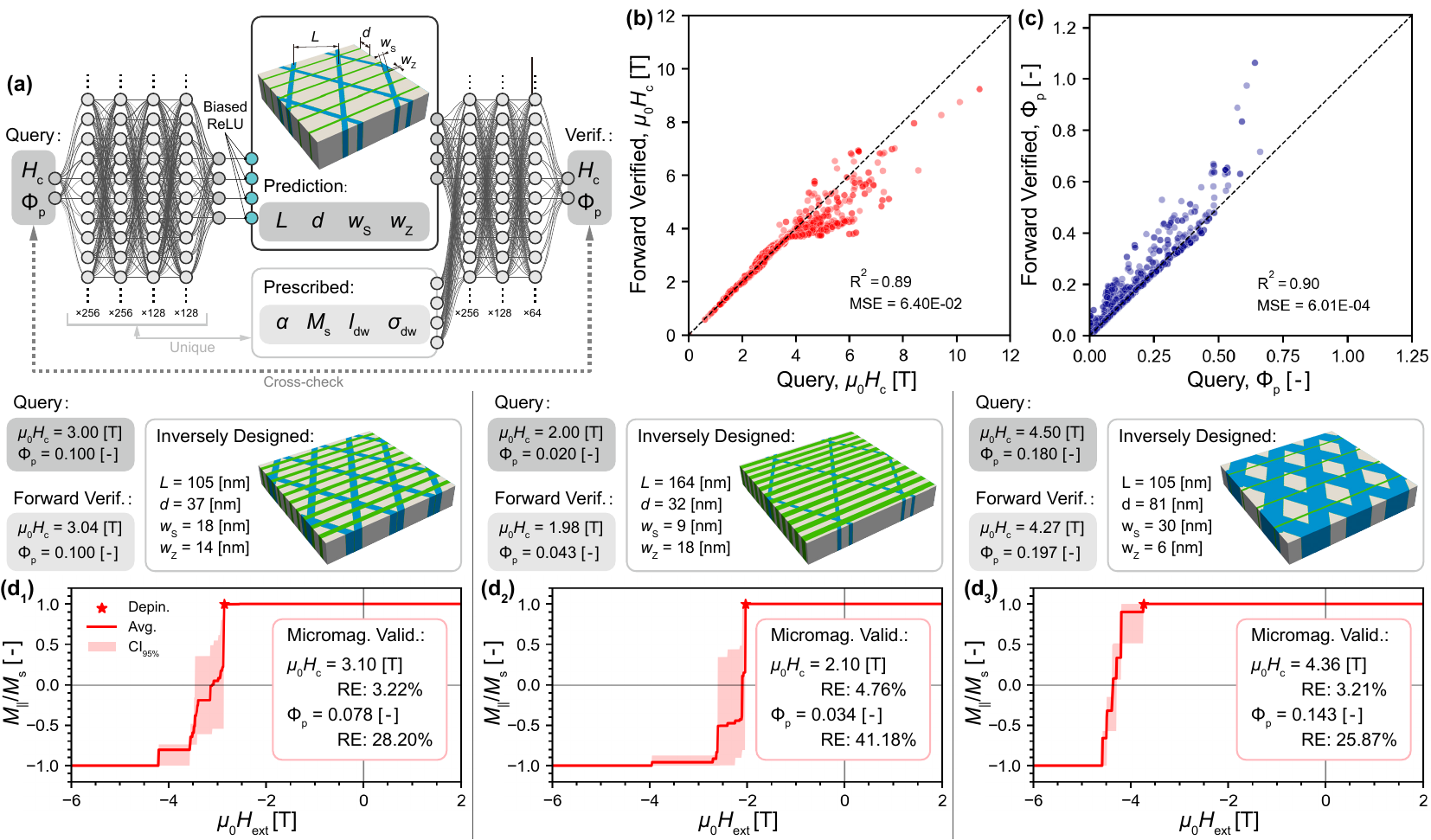}
	\caption{\small\textbf{Inverse design of $\smco$ nanostructures
    }. (a) Schematic of the inverse design process. The inverse model ($\iNN$) reads a set of queried values $\Hc$ as input and $\Pp$ and feeds back a set of nanostructural descriptors ($L$, $d$, $\ws$, and $\wz$) as outputs, which are then fed into the forward model ($\fNN$) for verification. Micromagnetic descriptors ($\alpha$, $\Ms$, $\ldw$, and $\sigdw$), on the other hand, are prescribed during the training of the $\iNN$ and inputs of the $\fNN$ for verification. Parity plots of (b) $\Hc$ and (c) $\Phi$ on inversely designed nanostructures vs. queried values. (d$_1$)-(d$_3$) Three inversely designed nanostructures based on queried $\Hc$ and $\Pp$, respectively. The demagnetization curves from micromagnetic simulations, used for validation, are demonstrated.    }
	\label{fig5}
\end{figure}

In \figref{fig5}, we demonstrate the data-driven inverse design of $\smco$ nanostructure from queried $\Hc$ and $\Pp$. In contrast to the $\fNN$ model elaborated above, inverse design directly delivers the design parameters (nanostructural and micromagnetic descriptors) for the specified target properties ($\Hc$ and $\Pp$). This differs from the optimization task, which aims to find extreme values of the properties while scanning the design space. Though also interesting, we leave it for near-future work. For the inverse design approach, another NN model, denoted by $\iNN$, was trained mapping $\Hc$ and $\Pp$ onto the design parameters, as depicted in \figref{fig5}a. Note that it makes use of the forward $\fNN$ model in the loss function of $\iNN$ to improve the accuracy and avoid multiplicity of inverse predictions. Details are summarized in section \textit{Method}.  
At the end, $\fNN$ was also employed to cross-check the corresponding property value of the design parameters suggested by the inverse model. 

As shown in \subsfigref{fig5}{b}{c}, the queried coercivity $\Hc$ and inhomogeneity factor $\Pp$ are forward verified using the $\fNN$ model on inversely designed nanostructures. Here, the micromagnetic descriptors are prescribed as $\alpha=0$. $\Ms$, $\sigdw$, and $\ldw$ take values of \ce{SmCo5} as 1:5-phase, \ce{Sm2Co_{17}} as 2:17-phase, and Z-phase from literature (\tabref{tab:1}). Meanwhile, to improve the training, 10,000 points are generated under the prescribed micromagnetic descriptors by $\fNN$ model as the training data for the $\iNN$ model. This means each $\iNN$ model is unique to the prescribed micromagnetic descriptors. Consequently, the nanostructural descriptors $L$, $d$, $\ws$, and $\wz$ are the sole outputs of the $\iNN$. Verification is performed by feeding the inversely designed descriptors into the $\fNN$ model. For $\Hc$, the scatter points cluster tightly along the 45$\deg$ diagonal in the low-value region ($\mu_0\Hc<4$~T), indicating that the inverse predictions of the $\iNN$ are highly consistent with the queried values. In the high-value region ($\mu_0\Hc>4$~T), however, the scatter distribution shows increasingly larger deviations, with a systematic tendency toward underestimation. This behavior yields an overall $R^2=0.89$ and $\text{MSE}=6.40\times10^{-2}$, showing slightly lower accuracy compared to $\Pp$. Similar to the $\fNN$ results. This underestimation is likely inherited from the $\fNN$ model, as it exhibits the same issue when cross-checked with the micromagnetic simulation results during training. For $\Pp$, the inverse predictions achieve $R^2=0.90$ and $\text{MSE}=6.01\times10^{-4}$, with scatter points closely following the diagonal but showing a mild systematic overestimation across the entire range. This overestimation trend will also be evident in the subsequent micromagnetic validation.

Three cases of nanostructural inverse design are demonstrated \subssfigref{fig5}{d1}{d3}, where the $\Hc$ and $\Pp$ of the inversely designed nanostructures are validated by micromagnetic simulations.  Across the three cases, the resulting $\Hc$ values agree well with the queried values, with relative errors of less than $5\%$. Notably, the $\iNN$ occasionally returns nanostructures with parameters outside the range of the training dataset ($L\in[30,150]$, $d\in[10,55]$, $\ws\in[3,15]$, $\wz\in[2,10]$~nm), suggesting a certain degree of extrapolative capability. For $\Pp$, however, micromagnetic validation reveals a tendency toward overestimation, with relative errors exceeding $25\%$. Nevertheless, the inverse design approach performs robustly within the stable range of queried $\Hc$ (particularly in the low-value regime, as shown in \subfigref{fig5}{b}), where the $\iNN$ yields nanostructures with minimal relative error. This outcome further motivates the extension of the dataset toward the high-coercivity regime in future work.

\subsection*{Angular dependence of the coercivity with pinning contribution}

\begin{figure}[!t]
	\centering
 	\includegraphics[width=18cm]{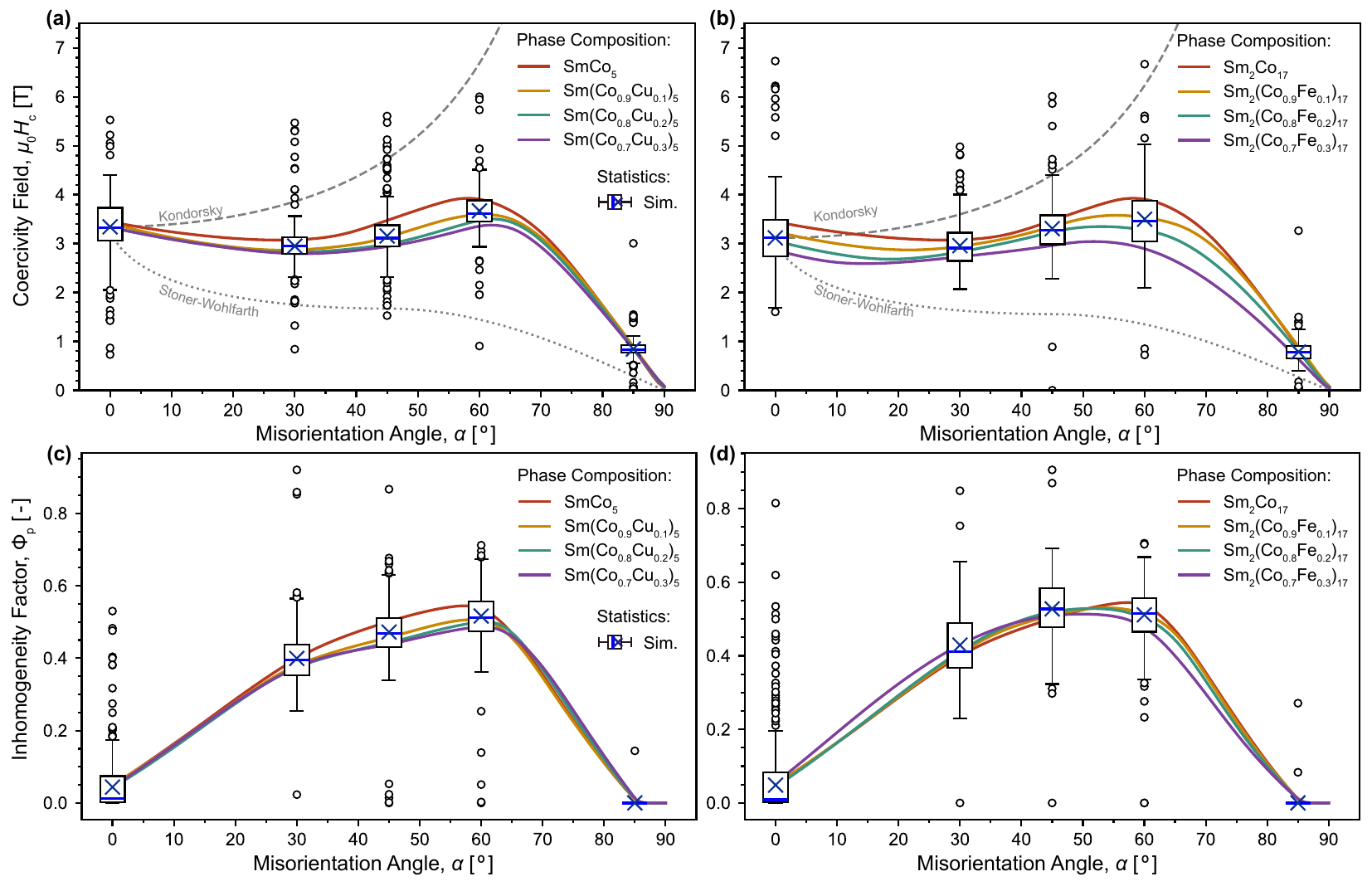}
	\caption{\small\textbf{Angular dependence of the coercivity and inhomogeneity factor}. Variation of coercivity $\Hc$ with misorientation angle $\alpha$ for (a) different Cu contents in the 1:5-phase and (b) different Fe contents in the 2:17-phase. The corresponding inhomogeneity factor $\Pp$ is shown for the same conditions (c) different Cu contents in the 1:5-phase and (d) different Fe contents in the 2:17-phase. Statistical distributions of the micromagnetic simulation results are represented as boxplots, with the mean denoted by the cross and outliers indicated by dots. For clarity, only 5\% of the outliers are shown.
}
	\label{fig6}
\end{figure}

The angular dependence of coercivity $\Hc(\alpha)$ obtained from $\fNN$ prediction is shown in \subsfigref{fig6}{a}{b} for varying Cu content in the 1:5-phase and Fe content in the 2:17-phase, respectively. Here, results from high-throughput micromagnetic simulations are illustrated as boxplots, demonstrating the statistical descriptors such as the mean (cross), the median, quartiles, and the outliers (dots). Theoretical models for the angular dependence of $\Hc$, namely the Stoner–Wohlfarth model and the Kondorsky model, are also denoted as dashed and dotted lines, respectively. Taking the simplification that nucleation field $\Hn$ and depinning field $\Hp$ are all merged as one $\Hc$ at $\alpha=0\deg$ (noted as $H_\mathrm{c,0}$), the $\Hc$ at varying $\alpha$ is than described by $\Hc(\alpha)=H_\mathrm{c,0}f(\alpha)$ with the angular dependence $f(\alpha)$ of two models as
\begin{equation}
    f_\mathrm{SW}(\alpha) = \begin{cases}
\displaystyle {\left[ \left(\cos \alpha \right)^{2/3} + \left(\sin \alpha \right)^{2/3} \right]^{-3/2}}, 
& 0 \leq \alpha \leq \tfrac{\pi}{4}, \\[2ex]
\displaystyle \frac{\sin(2\alpha)}{2}, 
& \tfrac{\pi}{4} \leq \alpha \leq \tfrac{\pi}{2}.
\end{cases},\quad 
f_\mathrm{K}(\alpha) = \frac{1}{\cos(\alpha)}.
\end{equation}
It is worth noting that the Stoner-Wohlfarth model views magnetization reversal as a coherent rotation. Since coherent rotation also characterizes the formation of the nucleation \cite{bance2014a}, the material with angular dependence close to $f_\mathrm{SW}$ is regarded as the nucleation-type. Meanwhile, the Kondorsky model is derived based on the pinning mechanisms, one with angular dependence close to $f_\mathrm{K}$ is thus known as the pinning-type. 

For varying Cu content in the 1:5-phase (\subfigref{fig6}{a}), $\Hc$ first decreases, reaching a local minimum near $30^\circ$, then gradually rises to a local maximum around $60^\circ$, followed by a sharp drop to zero at $90^\circ$. Increasing Cu content reduces $\Hc$ without altering the angular dependence; the curves shift downward without changing their overall angular profile. The statistical distributions (boxplots) of micromagnetic simulations with different Cu contents largely overlap with the $\fNN$ predictions, demonstrating that composition primarily shifts the coercivity value rather than reshaping its angular dependence.
For varying Fe contents in the 2:17-phase (\subfigref{fig6}{b}), $\Hc$ is systematically lower than that in the 1:5-phase. With increasing Fe contents, further reduced $\Hc$ is observed, owing to its effect on decreasing $\Ku$ of the 2:17-phase (matrix). Meanwhile, downshifting of the local maximum and local minimum of $\Hc$ is evident with increasing Fe, i.e., local minimum around $30\deg$ with $\ce{Sm2Co17}$ shifts to $10\deg$ with $\ce{Sm2(Co_{0.7}Fe_{0.3})17}$, and local maximum around $60\deg$ with $\ce{Sm2Co17}$ shifts to $52\deg$ with $\ce{Sm2(Co_{0.7}Fe_{0.3})17}$. 
The overall orientation dependence remains almost unchanged with variations in Cu/Fe contents, consistently falling between the idealized Stoner–Wohlfarth and Kondorsky models. This suggests that magnetization reversal is not controlled solely by nucleation but also involves defect-assisted nucleation and domain-wall motion. The boxplot statistics support this interpretation: the widest distributions and largest scatter occur at intermediate orientations ($30\deg$–$60\deg$), where pinning and nucleation introduce variability, while the narrower spreads at $0\deg$ and $90\deg$ are consistent with the coherent-rotation behavior predicted by the Stoner–Wohlfarth model.

From another perspective, \subsfigref{fig6}{c}{d} illustrates the orientation dependence of $\Pp$, which by definition (\eqref{eq:Pp}) quantifies the fraction of $\Hc$ arising from pinning effects. For both increasing Cu contents in the 1:5-phase and increasing Fe contents in the 2:17-phase, $\Pp$ rises from near 0.05 at $0^\circ$, peaks around $50^\circ$–$60^\circ$, and then rapidly drops to zero at $90^\circ$. The maximum $\Pp$ occurs in the same angular range where $\Hc$ exhibits a local maximum. This supports the aspect that the enhanced $\Hc$ around $55^\circ$–$65^\circ$ is largely due to pinning effects, while the immediate switching at $0^\circ$ and coherent rotation at $90^\circ$ suppress pinning contributions. Increasing Cu/Fe contents primarily reduces the overall level of $\Pp$. Notably, for higher Fe contents in the 2:17-phase, the peak of $\Pp$ shifts from $60^\circ$ with \ce{Sm2Co17} to about $52^\circ$ with \ce{Sm2(Co_{0.7}Fe_{0.3})17}, consistent with the concurrent shift of the $\Hc$ peak.

Overall, the angular dependence of $\Hc$ and $\Pp$ reflects a typical misorientation-associated domain-wall pinning, as observed in realistic permanent magnets \cite{panagiotopoulos2004, jahn1987}. Variations in Cu/Fe contents primarily shift the magnitude of $\Hc$, while indirectly influencing its angular dependence through changes in $\Pp$, though the overall trend remains essentially unchanged.

\subsection*{Nanostructural dependence of the coercivity with pinning contribution}

\begin{figure}[!t]
	\centering
 	\includegraphics[width=18cm]{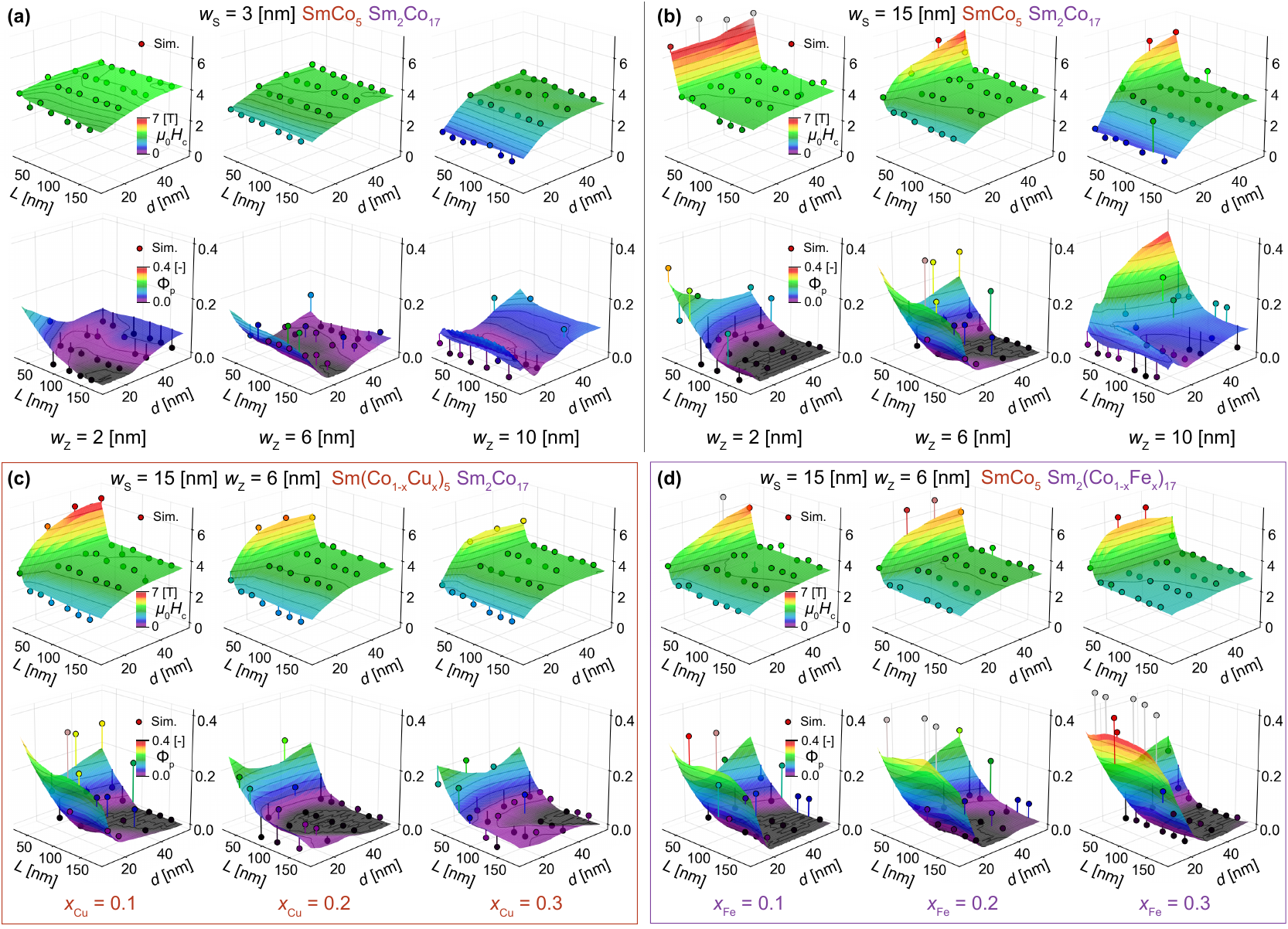}
	\caption{\small\textbf{Nanostructural dependence of the coercivity and inhomogeneity factor}. Contour surface of $\Hc$ and $\Pp$ vs. characteristic distance ($L$ and $d$) at various thickness of \ce{SmCo5} and \ce{Sm2Co17} phases: (a) $\ws=3$ nm, and (b) $\ws=15$ nm, with $\wz=2,6,10$, respectively. Contour surfaces are also illustrated for various Cu/Fe contents: (c) \ce{Sm(Co_{1-x}Cu_{x})5} ($x_\mathrm{Cu}=0.1,0.2,0.3$) and \ce{Sm2Co17} phases, and (d) \ce{SmCo5} and \ce{Sm2(Co_{1-x}Fe_{x}17} ($x_\mathrm{Fe}=0.1,0.2,0.3$) phases, with fixed $\ws=15$ nm and $\wz=6$ nm.
}
	\label{fig7}
\end{figure}

\subsfigref{fig7}{a}{b} presents how nanostructural descriptors, i.e., $L$, $d$, $\ws$ and $\wz$, influence $\Hc$ and $\Pp$ for the case of $\alpha=0\deg$. Here, both 1:5- and 2:17-phases are without Cu/Fe contents, respectively. The corresponding contour surfaces, predicted by $\fNN$ models, present the trend of $\Hc$ and $\Pp$ on nanostructural descriptors, while dots indicate micromagnetic results. In the top row, a general trend of $\Hc$ on nanostructural descriptors can be observed, i.e., for given $\ws$ and $\wz$, increasing $d$ and decreasing $L$ lead to the improvement of $\Hc$. An extended contour plot is presented in Supplementary Fig. \hl{4}.  
When $\ws = (\ws)_\mathrm{min} = 3~\si{nm}$ and $\wz = (\wz)_\mathrm{min} = 2~\si{nm}$, the dependence of $\Hc$ on $L$ and $d$ is relatively weak. Nevertheless, a general trend can be observed -- reducing $L$ and $d$ tends to improve $\Hc$. This effect becomes increasingly pronounced as $\ws$ and $\wz$ increase. Notably, a sharp rise in $\Hc$ emerges in the range $L \in [30, 50]~\si{nm}$ as $\ws$ approaches $(\ws)_\mathrm{max} = 15~\si{nm}$. Similarly, a rapid decline in $\Hc$ is evident in the range $d \in [10, 15]~\si{nm}$ as $\wz$ increases toward $(\wz)_\mathrm{max} = 10~\si{nm}$.
Recalling the anisotropic field $\Hani$ of each nanoscopic phase in \tabref{tab:1} and \subfigref{fig3}{d}, the 1:5-phase has the highest $\Hani$ among the three phases, while the 2:17-phase has the lowest one. We further consider two normalized characteristic quantities
\begin{equation}
    \vfs=\frac{\ws\left(2 L-\ws\right)}{L^2}-\frac{\ws\wz}{Ld},\quad\vfz=\frac{\wz}{d},
    \label{eq:vf}
\end{equation}
 adapting physical meaning as the volumetric fraction of 1:5- and Z-phases in an ideal periodic microstructure. In this regard, the nanostructure with large $\ws$ and small $L$ suggests a relatively high volume fraction of 1:5-phase with the highest $\Hani$ among three phases, leading to the drastic rise of $\Hc$. On the other hand, the bottom row of \subsfigref{fig7}{a}{b} depicts the variation of $\Pp$ w.r.t. nanostructural descriptors. The corresponding extended contour plot is also presented in Supplementary Fig. \hl{11}. 
When both $\ws$ and $\wz$ are small, e.g., $\ws = (\ws)_\mathrm{min} = 3~\si{nm}$ and $\wz = (\wz)_\mathrm{min} = 2~\si{nm}$, the $\Pp$ contour surface remains near zero across most of the contour surface. Nonetheless, a clear trend can be evident -- decreasing either $d$ or $L$ improves $\Pp$, implying enhanced pinning effects. Under identical $\ws$ and $\wz$, reducing $d$ is more effective than reducing $L$ in increasing $\Pp$. Even when $\ws > \wz$ (e.g., $\ws = 15~\si{nm}$ and $\wz = 6~\si{nm}$), the increase in $\Pp$ within $d \in [10,15]~\si{nm}$ is larger than that within $L \in [30,50]~\si{nm}$. This suggests that the Z-phase acts as a more effective pinning site provider than the 1:5-phase, while the 1:5-phase primarily enhances $\Hc$ through its higher anisotropy field $\Hani$. Moreover, decreasing $d$ not only increases the number of Z-2:17 interfaces but also introduces additional Z-1:5 interfaces as well as Z-1:5-2:17 three-phase junctions, potentially strengthening the total pinning contribution by increasing the number of pinning sites. A notable ridge in $\Pp$ is observed around $d \in [10,14]~\si{nm}$ when $\wz$ approaches $10~\si{nm}$, attributed to the structure being filled by the Z-phase ($\wz = 10~\si{nm}$), which removes pinning sites and leads to $\Pp \to 0$.

\subsfigref{fig7}{c}{d} illustrate the nanostructure dependence of $\Hc$ and $\Pp$ as the Cu/Fe contents are varied in the 1:5- and 2:17-phases, respectively. To facilitate a discussion with variable control, we separately increase the Cu content in the 1:5-phase ($\ce{Sm(Co_{1-x}Cu_{x})_5}$ with $x_\mathrm{Cu}$ from 0.1 to 0.3) while keeping the 2:17-phase ($\ce{Sm2Co_{17}}$) unchanged (\subfigref{fig7}{c}), and the Fe content in the 2:17-phase ($\ce{Sm2(Co_{1-x}Fe_{x})_{17}}$ with $x_\mathrm{Fe}$ from 0.1 to 0.3) while holding the 1:5-phase ($\ce{SmCo_{5}}$) constant (\subfigref{fig7}{d}). Both $\ws$ and $\wz$ are also fixed at 15 and 6~nm, respectively. Expanded contour plots are available in Supplementary Fig.~\hl{5}-\hl{10} for $\Hc$, and Supplementary Fig.~\hl{12}-\hl{17} for $\Pp$. With increasing $x_\mathrm{Cu}$, a gradual decrease in the peak value of $\Hc$ is evident in the $L \in [30,50]~\si{nm}$ range, which can be attributed to the reduction of $\Hani$ in the 1:5-phase as shown in \subfigref{fig3}{d} and \tabref{tab:1}. Additionally, $\Pp$ tends to decrease at lower $L$ and $d$ values as $x_\mathrm{Cu}$ increases, with a particularly noticeable reduction of the peak within the $d \in [10,15]~\si{nm}$ region, suggesting a diminished pinning effect. This phenomenon is likely due to the reduced contrast in $\sigdw$ between the 1:5-, 2:17-, and Z-phases as $x_\mathrm{Cu}$ increases, which decreases the energy barrier for domain-wall penetration at phase junctions and diminishes the strength of pinning sites. In contrast, increasing $x_\mathrm{Fe}$ in the 2:17-phase has a relatively minor influence on the nanostructure dependence of $\Hc$ and $\Pp$, mainly because the reduction in $\sigdw(x_\mathrm{Fe})$ within the 2:17-phase is less pronounced compared to that caused by $x_\mathrm{Cu}$ in the 1:5-phase. Specifically, $\sigdw$ in 2:17-phase decreases merely 10\% for $x_\mathrm{Fe}$ increasing from 0 to 0.3, while $\sigdw$ in 2:17-phase drops almost 32\% for $x_\mathrm{Cu}$ increasing from 0 to 0.3. Nevertheless, when $x_\mathrm{Fe}$ reaches 0.3, the peak value of $\Pp$ in the $d \in [10,15]~\si{nm}$ range rises, which can be attributed to enhanced Z-1:5-2:17 three-phase junctions resulting from increased contrast in $\sigdw$.

\begin{figure}[!t]
	\centering
 	\includegraphics[width=18cm]{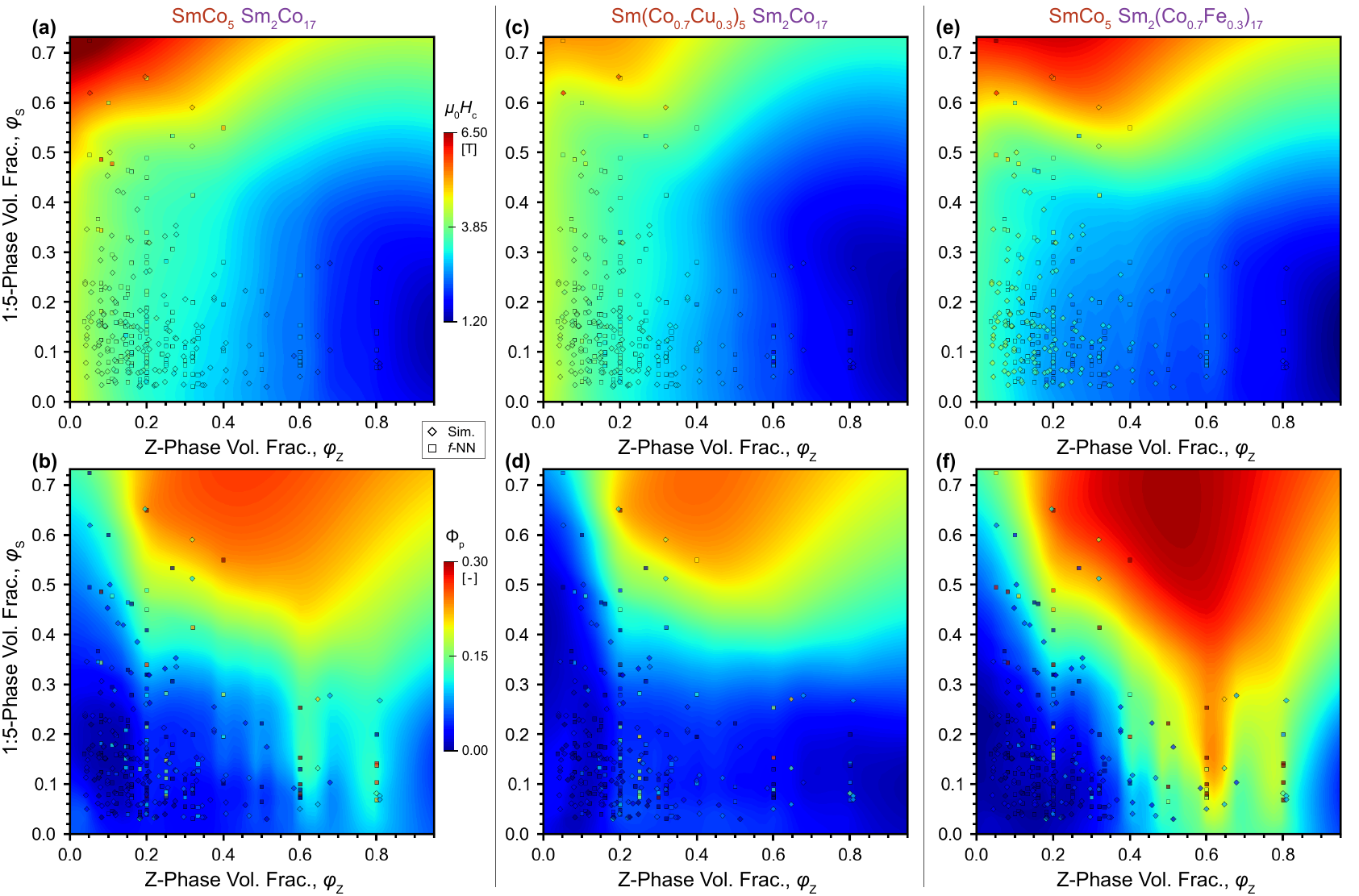}
	\caption{\small\textbf{Phase fraction dependence of the coercivity and inhomogeneity factor}. Contour surfaces of $\Hc$ and $\Pp$ vs.
 volume fractions $\vfs$ and $\vfz$ of the corresponding phases with varying Cu/Fe contents: (a) $\Hc$ and (b) $\Pp$ with \ce{SmCo5} and \ce{Sm2Co17} phases, (c) $\Hc$ and (b) $\Pp$ with \ce{Sm(Co_{0.7}Cu_{0.3})5} and \ce{Sm2Co17} phases, and (d) $\Hc$ and (e) $\Pp$ with \ce{SmCo5} and \ce{Sm2(Co_{0.7}Fe_{0.3}17} phases.
}
	\label{fig8}
\end{figure}

Recalling \eqref{eq:vf}, we further examine the trends of $\Hc$ and $\Pp$ on the volume fraction of the nanoscopic phases, which is presented in \figref{fig8}. The contour surface is given by the regression of a thin-plate spline (TPS) surface on a dataset containing 10,000 predicted data points by the $\fNN$ model. We also sample 150 data points each by simulation and prediction, and illustrate them together with the contour surface. In \subsfigref{fig8}{a}{b}, $\Hc$ and $\Pp$ are illustrated versus the volume fractions $\vfs$ of \ce{SmCo5} (1:5-phase, $x_\mathrm{Cu}=0$) and $\vfz$ of \ce{Sm2Co17} (2:17-phase, $x_\mathrm{Fe}=0$). High $\Hc$ appears in the top-left region with large $\vfs$ and low $\vfz$, consistent with the dominant role of the 1:5-phase, which possesses the highest $\Hani$, in enhancing coercivity. The maximum $\Hc$ occurs at about $\vfs \sim 0.7$ with nearly no Z-phase, suggesting that a sufficient 1:5-phase fraction is essential for maximizing coercivity. By contrast, $\Pp$ peaks in a different regime, around $\vfs \sim 0.7$ and $\vfz \sim 0.4$–$0.5$, where the coexistence of 1:5- and Z-phases increases the number of Z-1:5 interfaces and Z-1:5-2:17 three-phase junctions, thus strengthening the pinning effect. 

When considering the addition of Cu/Fe in the 1:5-/2:17-phases, a peak of $\Hc$ emerges along varying $\vfz$. In \subsfigref{fig8}{c}{e}, the maximum $\Hc$ occurs around $\vfz \sim 0.2$–$0.25$ for \ce{Sm(Co_{0.7}Cu_{0.3})_5} as the 1:5-phase with \ce{Sm_2Co_{17}} as the 2:17-phase, and $\vfz \sim 0.2$–$0.55$ for \ce{Sm2(Co_{0.7}Fe_{0.3})_{17}} as the 2:17-phase with \ce{SmCo5} as the 1:5-phase. Notably, although the region of peak $\Pp$ (at $\vfz \sim 0.4$–$0.5$) remains largely unchanged, it does not fully overlap with the region of peak $\Hc$, suggesting that Cu/Fe additions can moderate the coercivity of the nanostructure in a non-linear manner, rather than simply scaling the $\Hani$ of the corresponding phases or enhancing the pinning effect. Overall, these trends indicate that while coercivity generally benefits from increasing the 1:5-phase fraction, an optimized coercivity arises from a balanced mixture of 1:5- and Z-phases, implying that the optimal microstructure requires jointly tuning $\vfs$ and $\vfz$ rather than maximizing a single phase fraction.




\subsection*{Sensitivity analyses on geometric and magnetic properties of nanoscopic phases}

\begin{figure}[!t]
	\centering
 	\includegraphics[width=18cm]{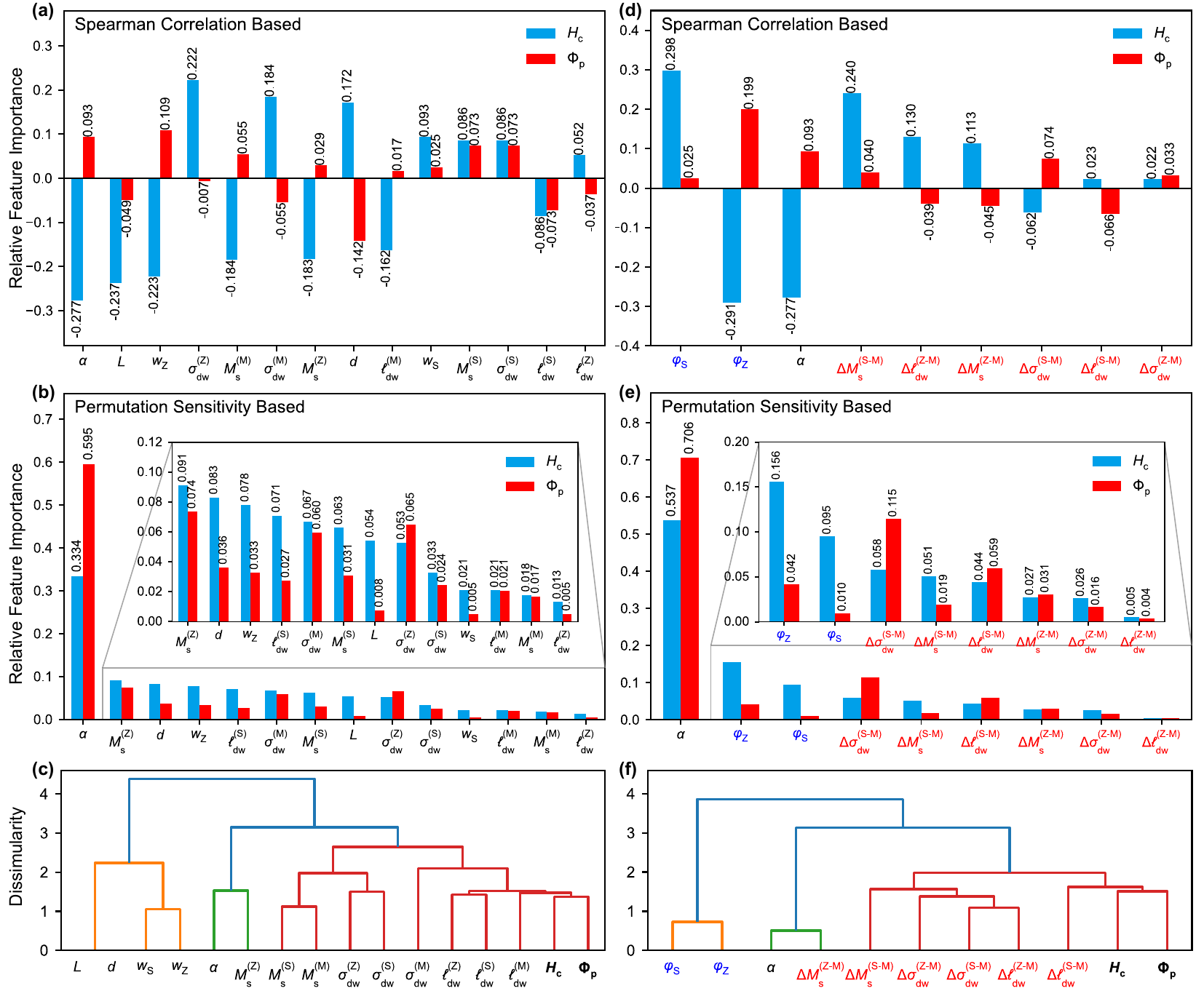}
	\caption{\small\textbf{Sensitivity analysis on nanostructural and micromagnetic descriptors}. Barplots of relative feature importance of nanostructural and micromagnetic descriptors for $\Hc$ and $\Pp$ based on (a) Spearman correlation and (b) permutation test. (c) Hierarchical clustering of these quantities. Superscripts `(M)', `(S)', and `(Z)' denote the cellular 2:17-phase (matrix), strip-shaped 1:5-phase, and platelet-shaped Z-phase, respectively. For comparison, feature importance using micromagnetic descriptor contrasts (red labels) and phase fraction (blue labels) is shown based on (e) Spearman correlation and (f) permutation test, with corresponding hierarchical clustering in (g).}
	\label{fig9}
\end{figure}

Sensitivity analyses based on Spearman's correlation (\figref{fig9}{a}) and permutation sensitivity test (\figref{fig9}{b}) importance were conducted to identify the nanostructural and micromagnetic descriptors most relevant to $\Hc$ and $\Pp$. Therefore, here we provide an integrated discussion of the effects of each phase and its descriptors by combining the results of both feature-importance analyses. 

First of all, the misorientation angle $\alpha$, which determines whether magnetization reversal proceeds via rotation or domain-wall motion, emerges as the dominant factor, imposing the strongest negative correlation with $\Hc$ and at the same time being the most effective descriptor for $\Pp$. This highlights the key role of magnetocrystalline misorientation in determining both the coercivity and the extent of pinning. Structural parameters provide the next level of influence: $L$ and $\wz$ are negatively correlated with $\Hc$, whereas $d$ and $\ws$ have weaker but positive effects. Importantly, $d$ and $\wz$ also rank among the top descriptors for $\Pp$, confirming that the geometry of the Z-phase plays a significant role in pinning.

The micromagnetic descriptors of the Z-phase further suggest their role in enhancing the resistance to the magnetization reversal. Both $\Msz$ and $\sigdwz$ correlate positively with $\Hc$, and their high permutation sensitivity suggests that Z-phase saturation magnetization and domain-wall energy act as effective barriers to reversal. For $\Pp$, $\sigdwz$ and $\sigdwm$ carry considerable importance, implying that pinning is governed not only by Z-phase geometry but also by the domain-wall energy contrast between Z- and 2:17-phases. By comparison, the micromagnetic properties of the 1:5-phase ($\Mss$, $\sigdws$, and $\ldws$) only weakly correlated to $\Hc$, but achieve intermediate importance for $\Pp$, ranking just after $\alpha$. This indicates that while the 1:5-phase has a relatively moderate impact on coercivity, its local properties still affect inhomogeneity. 

Finally, the 2:17-phase micromagnetic descriptors ($\Msm$, $\sigdwm$, and $\ldwm$) demonstrate intermediate contributions to $\Hc$ and relatively weak contributions to $\Pp$, indirectly reflecting its role as a matrix phase providing continuity rather than strong pinning sites. Overall, these combined results suggest a clear division of roles: coercivity is primarily governed by misorientation and reinforced by Z-phase micromagnetic properties, whereas pinning contribution is dominated by Z-phase geometry, with the 1:5-phase offering a weaker but non-negligible effect.

It is worth noting that using the direct nanostructural and micromagnetic descriptors of the phases does not always capture their explicit effects on coercivity and underlying pinning contribution. As elaborated earlier in this work, the pinning event originates from the energy perturbations induced when a domain wall penetrates nanoscopic phases with differing magnetic properties. Meanwhile, the overall amount of these phases further influences coercivity by contributing to the effective $\Hani$ of the material (like 1:5-phase). Therefore, it is necessary to perform sensitivity analyses not only directly on the descriptors but also on their relative values between magnetic properties of phases, as well as on the phase fractions (\eqref{eq:vf}). In this context, micromagnetic descriptor contrasts between the matrix and the embedded phases are introduced as
\begin{equation}
    \begin{split}
        \Delta\Mssm=\Mss-\Msm,&\quad
        \Delta\Mszm=\Msz-\Msm,\quad\\
        \Delta\sigdwsm=\sigdws-\sigdwm,&\quad
        \Delta\sigdwzm=\sigdwz-\sigdwm,\quad\\
        \Delta\ldwsm=\ldws -\ldwm,& \quad
        \Delta\ldwzm=\ldwz -\ldwm.
    \end{split}
\end{equation}
According to \tabref{tab:1}, both $\Delta \Ms$ are negative and both $\Delta \ldw$ are positive within the parameter space, while $\Delta \sigdw^{(S-M)}$ is positive and $\Delta \sigdw^{(Z-M)}$ is negative. This indicates that the 1:5–2:17 interface acts as a repulsive pinning site, whereas the Z–2:17 interface serves as an attractive one.

\subsfigref{fig9}{d}{e} presents the sensitivity analyses on these contrast quantities and the phase fractions with respect to $\Hc$ and $\Pp$. In agreement with \subsfigref{fig9}{a}{b}, the misorientation angle $\alpha$ remains an effective factor, consistently imposing a strong influence on both $\Hc$ and $\Pp$. The usage of phase fractions further supports the earlier interpretation: $\vfs$ shows a strong positive effect on $\Hc$, which is coherent with the contours presented in \figref{fig8}. Meanwhile, $\vfz$ correlates negatively with $\Hc$ while contributing significantly to $\Pp$, reinforcing the role of the Z-phase as the primary provider of pinning. The introduction of contrast quantities offers a more direct assessment of the pinning mechanism. Notably, the influence of the 1:5-phase becomes more evident when evaluating contrast quantities to the 2:17-phase. In particular, permutation sensitivity analysis highlights its effect on $\Hc$ only next to $\alpha$ and the phase fractions. At the same time, $\Delta\sigdw^{(S-M)}$ emerges as the most influential factor for $\Pp$ after $\alpha$, suggesting that the contribution of the 1:5-phase to the pinning mechanism cannot be overseen.

Based on Spearman’s correlations, the effects of the descriptors are hierarchically clustered in \figref{fig9}{c}. The cluster arrangement indicates feature similarity, with branch point heights reflecting dissimilarity. Here $L$, $d$, $\ws$, and $\wz$ form a closely related group (orange), consistent with their role in describing the geometry of the nanoscopic phases. A second cluster links $\alpha$ with $M_s^{(Z)}$, reflecting their shared impact on magnetization reversal (green). The remaining micromagnetic parameters, together with $\Hc$ and $\Pp$, form the most significant cluster (red), suggesting that these features are more strongly associated with coercivity and underlying pinning mechanisms. A similar trend is obtained when phase fractions and contrast quantities are analyzed (\figref{fig9}{f}), where $\vfs$ and $\vfz$ cluster as nanostructural descriptors, while the contrast quantities group with $\Hc$ and $\Pp$, confirming consistency with the previous analysis and underscoring the added relevance of relative descriptors.

\section{Discussion}

In this work, we integrate machine learning approaches and high-throughput micromagnetic simulation to investigate the influence of the cellular nanostructure on the coercivity of the $\smco$ permanent magnets. After conducting data-driven forward analyses, we found:
\begin{enumerate}[(i)]
    \item Among nanostructural descriptors, both $\Hc$ and $\Pp$ are most sensitive to the Z-phase geometry, characterized by the interval $d$ and the thickness $\wz$. Besides, $L$ and $\wz$ are negatively correlated to $\Hc$, whereas $d$ and $\ws$ provide weaker positive correlation. Increasing $\vfs$ (1:5 fraction) generally enhances $\Hc$, consistent with the role of the 1:5-phase as the provider of the highest $\Hani$, whereas increasing $\vfz$ (Z-phase fraction) promotes $\Pp$ by providing more effective pinning sites, which is coherent to the finding that both $\Hc$ and $\Pp$ are strongly sensitive to $\sigdwz$ and $\Msz$. The highest $\Pp$ is achieved in microstructures with about 40-50\% Z-phase, where the abundance of Z-1:5 interfaces and Z-1:5-2:17 junctions maximizes domain-wall energy fluctuations. However, excessive Z-phase content reduces $\Hc$ because of its lower intrinsic anisotropy. This establishes a clear division of roles: the 1:5-phase enhances coercivity by contributing high anisotropy, while the Z-phase strengthens pinning but simultaneously limits $\Hc$ when dominant. An optimal coercivity thus resides in a delicate balance between 1:5-phase and Z-phase.

    \item Effects of Cu and Fe contents are asymmetric. Incorporating Cu into the 1:5-phase effectively reduces $\Hc$ and weakens $\Pp$, mainly due to the decrease of $\Hani$ in the 1:5-phase and the diminished contrast in $\sigdw$ at phase boundaries. On the other hand, increasing Fe additions in the 2:17-phase merely slightly reduces $\Hc$, with $\Pp$ remaining stable. At higher Fe levels, however, $\Pp$ can increase locally within about 40-50\% Z-phase, primarily due to enhanced contrasts in magnetic properties between the matrix and inclusion phases (notably $\sigdwsm$ and $\Mssm$). These findings suggest that Cu in the 1:5-phase plays a significant role in moderating coercivity, while Fe in the 2:17-phase primarily fine-tunes the pinning effect without drastically altering coercivity. Moreover, since increasing Fe content in the 2:17-phase also improves the overall saturation magnetization of the magnet, adjusting the Fe composition may be another key optimization strategy. This strategy aligns with experimental findings \cite{hu2024a} and warrants further exploration.

    \item The misorientation angle $\alpha$ is the dominant factor across all descriptors, exerting the strongest negative correlation with $\Hc$ while also governing $\Pp$. Coercivity curves consistently fall between the Stoner-Wohlfarth and Kondorsky models, reflecting a mixed mechanism of nucleation, defect-assisted nucleation, and domain-wall motion. Pinning contributions ($\Pp$) peak at intermediate orientations ($50^\circ$-$60^\circ$), which is consistent with the local maxima of the coherence factor ($\Hc$). This indicates that grains with moderate misorientation benefit most from pinning effects. At $0^\circ$ and $90^\circ$, $\Pp$ vanishes, consistent with immediate switching and coherent rotation. Increasing the Cu/Fe content weakens the pinning effects (characterized by $\Pp$) overall. The Fe in the 2:17-phase further shifts the Pp peak toward smaller angles, resembling the shift observed in the Hc. This highlights the crucial roles of texture and misorientation in determining the magnitude of coercivity and the effective pinning contribution.
\end{enumerate}
Meanwhile, the presented work showcases the inverse design of nanostructural descriptors by querying $\Hc$ and $\Pp$ with acceptable precision. An interactive, web-based demonstration is available at \texttt{\url{https://smcotoolkit.streamlit.app}} for further examination. These ML-based models can readily find applications in various synthesis and engineering situations as a computationally cost-effective toolkit, for instance, the coercivity estimator and the coercivity-based nanostructure predictor. 

As a perspective, this work can be further conjugated with the scale-bridging hysteresis investigations, 
specifically, via connecting trained $\fNN$ with vector hysteron on a mesoscopic polycrystal \cite{yang2023}. By incorporating further nanostructure with distinct features (like monocrystalline-amorphous composite \cite{staab2024}) and properties (like the remanence and the maximum energy product \cite{gutfleisch2011}) in the extended dataset, it is also promising to accelerate the design of novel permanent magnets with smart and hierarchical microstructures and thus broad tailorable performances.


We plan to use the PyTorch deep learning framework for constructing and training these artificial neural network models. We will explore different combinations of hyperparameters to pinpoint models that exhibit satisfactory performance."


\section*{Method}

\subsection*{Electron microscopy}
	Electron-transparent specimens for TEM were fabricated using a Ga-focused ion beam (FIB) and plasma sputtering with dual-beam SEM/FIB systems (Zeiss Crossbeam 540 and ThermoFisher Helios G4 plasma FIB). Bright-field (BF) TEM imaging and selected-area electron diffraction (SAED) measurements were carried out in a conventional transmission electron microscope (JEOL JEM 2100F). High-resolution high-angle annular dark field scanning TEM (HAADF-STEM) imaging was carried out in an aberration-corrected system (JEOL JEM-ARM200F) operated at 200 kV.

 \subsection*{First principle calculation of micromagnetic parameters}
The magnetization saturation and exchange stiffness are evaluated by combining the DFT calculation and the atomistic spin dynamics (ASD) simulation. 
Using the post-processing code 'jx' in OpenMx~\cite{terasawa2019efficient, ozaki2003variationally} following a LDA + U self-consistent calculation~\cite{anisimov1997first}, the exchange coupling parameters $J_{ij}$ were calculated. In the self-consistent calculation, the magnitudes of U and J were selected as 6.7 and 0.7 eV, respectively, which are found to be suitable for most rare-earth elements. The energy cutoff and energy convergence criteria were set to 500 Ry and 1.0$\times$10$^{-8}$ Hartree, respectively. $10\times10\times2$ and $15\times15\times3$ k-meshes were used for self-consistent and $J_{ij}$ calculations, respectively. 
By using the calculated exchange parameters and by assuming a continuous long-wavelength function for spin distribution in the Heisenberg Hamiltonian~\cite{moreno2016temperature}, the exchange stiffness of \ce{(SmZr_2)Co_9} was evaluated using the Uppsala Atomistic Spin Dynamics (UppASD) software~\cite{eriksson2017atomistic}. The obtained magnetization saturation and exchange stiffness of \ce{(SmZr_2)Co_9} are 290 kA/m and 4.83 pJ/m, respectively. 
To validate the accuracy of this method, we also calculated the magnetization saturation and exchange stiffness of \ce{SmCo5} and \ce{Sm2Co_{17}}. For \ce{SmCo5} and \ce{Sm2Co_{17}}, the obtained $\Ms$ and $\Ku$ in this work are 780 \si{kA~m^{-1}} and 15.0 \si{pJ~m^{-1}}, 985 \si{kA~m^{-1}} and 19.48 \si{pJ~m^{-1}}, respectively, which agree well with the values reported in former works~\cite{katter1996new, liu2023a} and ones employed in this work. The relaxed lattice structures of three nanoscopic phases are presented in Supplementary Table \hl{2}.

\subsection*{High-throughput micromagnetic simulation}

In micromagnetics, magnetization is treated as a position-dependent vector $\mag(\pos)$, normalized with respect to the saturation magnetization, i.e., $\mag=\mathbf{M}/\Ms$. The free energy is formulated as a functional of $\mag(\pos)$ as
\begin{equation}
    \mathcal{F}=\int_V \left[f_\mathrm{e}+f_\mathrm{a}+f_\mathrm{ms}+f_\mathrm{zm}\right]\diff V,
    \label{eq:F}
\end{equation}
where the separate contributions $f_\mathrm{ex}$, $f_\mathrm{ani}$, $f_\mathrm{ms}$, and $f_\mathrm{zm}$ represent contributions from exchange, magneto-crystalline anisotropy, magnetostatics, and external magnetic field, respectively. They are formulated as
\begin{equation}
    \begin{split}
        f_\mathrm{e} &= \Ae \|\nabla\mathbf{m}\|^2, \\
        f_\mathrm{a} &= -\sum_i \Ku\left(\eau\cdot\mag\right)^{2i},\\
        f_\mathrm{ms} &= -\frac{1}{2}\mu_0\Ms\mag\cdot\H_\mathrm{dm},\\
        f_\mathrm{zm} &= -\mu_0\Ms\mag\cdot\Hvext.\\
    \end{split}
    \label{functionalContributions}
\end{equation}
Micromagnetic parameters ($\Ae$, $\Ku$, $\Ms$) for the nanoscopic phases are listed in \tabref{tab:1}. The magnetic exchange coupling on the interface between nanoscopic phases is formulated as
\begin{equation}
    \mathbf{H}_\mathrm{ex}^\mathrm{intf}=\frac{2S}{\mu_0}\langle{\Ae/\Ms}\rangle_\mathrm{H}\nabla\cdot\nabla\mathbf{m},
\end{equation}
where $\langle{\Ae/\Ms}\rangle_\mathrm{H}$ represents the harmonic mean of $\Ae/\Ms$ of adjacent bulk phases, and $S$ is a strength factor \cite{Vansteenkiste2014}. 

To remain consistent with the observed coherent interfaces (\figref{fig1}b) and to avoid additional complexity in the analyses (as $S<1$ introduces extra pinning effects on domain-wall migration \cite{leliaert2014current}), a default value of $S=1$ was adopted for all interfaces among the nanoscopic phases. Physically, this corresponds to the assumption of complete atomic contact at all interfaces \cite{hernando1992role}. To keep consistent with the observed coherent interfaces (\figref{fig1}b) and to avoid additional complexity (since $S<1$ introduces extra pinning effects on domain-wall migration \cite{leliaert2014current}), a default value of $S=1$ was adopted for all interfaces among the nanoscopic phases, corresponding to the assumption of complete atomic contact \cite{hernando1992role}.

The evolution of the magnetization configuration $\mv(\pos)$ under a cycling $\Hv_\mathrm{ext}$ is generally described by the Landau–Lifshitz–Gilbert (LLG) equation. Owing to the incomparable time scale of LLG-described magnetization dynamics (around nanoseconds) with respect to the one of hysteresis measurement (around seconds), constrained free-energy minimization of $\mathscr{F}$ is widely employed as an efficient alternative for evaluating hysteresis in permanent magnets \cite{Exl2014, schabes1988magnetization, furuya2015semi}. Based on the steepest conjugate gradient (SCG) method, the iteration scheme is given by
\begin{equation}
\begin{split}
&\frac{\mv^{(i+1)}-\mv^{(i)}}{\Delta^{(i)}}=\mv^{(i)}\times\frac{1}{\mu_0 \Ms}\left[\mv^{(i)} \times \varid{\fedf}{\mv^{(i)}} \right], \\
&\text{subject to}\quad|\mv|=1\label{eq:gov_mm}
\end{split}
\end{equation}
with iteration step size $\Delta^{(i)}$ and $\mv^{(i)}$ at step $i$. This scheme corresponds to the sole damping term of the LLG equation \cite{Exl2014, furuya2015semi}, implying quasi-static evaluation of hysteresis.

Boundary conditions (BCs) were applied depending on domain boundaries: periodic BC along the out-of-plane ($z$) direction using the macrogeometry approach \cite{fangohr2009new}, and Neumann BC elsewhere \cite{Vansteenkiste2014}. Since the in-plane configuration and domain-wall migration are primarily resolved, the grid number in the $z$-direction was reduced. Hence, two FD domains were used: $256 \times 256 \times 64~\si{nm}^3$ (S-domain) and $384 \times 384 \times 64~\si{nm}^3$ (L-domain) for cross-checking. It should be noted that sufficiently small finite-difference (FD) cells are required in micromagnetic simulations to adequately resolve both the fine microstructure details and the magnetic characteristic length (here, the domain-wall thickness). However, employing finer cells drastically increases the computational cost of each simulation with the same domain size, which in turn constrains the efficiency of high-throughput studies. To balance numerical accuracy and computational performance, we adopted a cell size of $0.5 \times 0.5 \times 2~\si{nm}^3$ for the high-throughput simulations after comparing the simulated coercivity $\Hc$ on two domains with different cell sizes (Supplementary Fig.~\hl{8}). Similarly, sampling in $\alpha$ for simulations with the L-domain is comparably sparser than those with the S-domain. In total, 10,800 out of 42,300 simulations (25.5\%) were employed on the L-domains.

More than 500 computational jobs were distributed in parallel across computing nodes equipped with the Nvidia$^\circledR$ Volta 100 GPU. Each distributed job batched 25–50 simulations in series to balance computation and queuing time. Processes including nanostructure generation, inputfile preparation, job submission and monitoring (including troubleshooting and subsequent inputfile correction/re-submission), as well as data collection and post-processing, were automated using customized Python scripts with the \texttt{multiprocessing} library.

\subsection*{Machine learning models for forward prediction}
In this work, an ML-based forward prediction model for micromagnetic simulations is essential for dealing with data recursion and extrapolation, particularly in the permutation sensitivity test and contour surface extraction. Different ML models were trained using the nanostructural/magnetic parameters as the input and corresponding POIs (here the coercivity $\Hc$ and inhomogeneity factor $\Pp$) from micromagnetic simulations as output, including the K-nearest neighbors, neural networks, gradient boosting, decision tree, random forest, and support vector machine. These models were implemented in customized Python scripts utilizing utilities from the \texttt{Scikit-learn} and \texttt{PyTorch} libraries \cite{pedregosa2011scikit, paszke2019pytorch}. Training of the forward models requires the minimization of the mean squared error ($\mathrm{MSE}$) between the true values of properties ($\Yv=\{\Hc,\Pp\}$) and the prediction, i.e., the loss function is as
 \begin{equation}
     \mathcal{L} = \min_{\wv} \frac{1}{n}\sum^n\|\fmod_{\wv}[\Xv]-\Yv\|^2,\label{eq:fmod}
 \end{equation}
where $\fmod$ represents aforementioned ML models with a set of model parameters $\wv$, $\Xv$ stands for the nanostructural/magnetic parameters $\Xv=\{L,d,\ws,\wz,\alpha, \Msz, \sigdwz, \ldwz\}$.  

Model performance was evaluated by the coefficient of determination ($\Rsq$) and MSE between the true (from micromagnetic simulations) and the predicted (by forward prediction models) values. They are correspondingly defined as
\begin{equation}
    \textrm{R}^2=\frac{\sum_{i}^N\left(\hat{Y}_i-\bar{Y}\right)^2}{\sum_{i}^N\left(Y_i-\bar{Y}\right)^2},\quad
    \mathrm{MSE}=\frac{1}{N} \sum_{i}^N\left(Y_i-\hat{Y}_i\right)^2,
\end{equation}
where $\hat{Y}_i$ is the predicted value by the forward prediction model, $Y_i$ is the test value from micromagnetic simulations, and $\bar{Y}$ is the mean of the test dataset with a size $N$.
The performance of the aforementioned six ML models was evaluated with their default hyperparameters, as presented in Supplementary Figs. \hl{2}-\hl{3}. To improve the performance of the forward prediction models, which were practically employed in the subsequent data-driven analyses, hyperparameter optimization was also conducted with the performance presented in \subsfigref{fig4}{a}{b}. As the NN model was eventually employed in this work for forward prediction (i.e., $\fNN$), the optimized hyperparameters are listed in \tabref{tab:m2}.

The selection of dataset size was examined in terms of the law of large number (LLN) and training performance. For any dataset with a size $N$, the dataset statistic descriptors (DSD) can be defined. Here the arithmetic means of $\Hc$ and $\Pp$ are employed as DSDs i.e., $\mu^{N}_{\Hc}=\frac{1}{N}\sum_i^{N}(\Hc)_i$ and $\mu^{N}_{\Pp}=\frac{1}{N}\sum_i^{N}(\Pp)_i$. According to the law of large number (LLN), $\mu^{N}_{\Hc}$ and $\mu^{N}_{\Pp}$ of any dataset with $N\rightarrow\infty$ should converge to a unified $\mu^{\infty}_{\Hc}$ and $\mu^{\infty}_{\Pp}$, which are regarded as the true DSDs of the dataset. Thus, the mean absolute percentage error (MAPE) can be defined to evaluate the descriptivity of the current dataset size, i.e.,
\begin{equation}
    \text{MAPE} = \frac{\left|\mu^{N}_{Y} -\mu^{\infty}_{Y}\right|}{\mu^{\infty}_{Y}}\times{100\%},
\end{equation}
where $Y=\Hc,\Pp$. In this work, the weak form of LLN, with $\mathrm{}_{\Hc}\leq1\%$, is taken as the convergence criterion, and the dataset size $N$ at that moment is used as the threshold dataset size for conducting data-driven and machine-learning approaches.

\begin{table}[]
\centering\small
\begin{threeparttable}
\caption{Optimized hyperparameters for the forward ($\fNN$) and inverse neural networks ($\iNN$).}
\begin{tabularx}{0.75\textwidth}{Xp{4cm}p{4cm}}
\hline
& $\fNN$ & $\iNN$    
\\ \hline 
Batch size &
128 &
128 \\
Input, (hidden), output layers
& 8\tnote{*}, (256, 128, 64), 2
& 2, (256, 256, 128, 128), 4    
\\ 
Activation function &
ReLU & ReLU \\
Optimizer &
Adam & Adam \\
Learning rate & 0.001 & 0.0001 \\
Epochs (stop/total) & 247/2000 & 100/100\tnote{**} \\
Numerical scaling & $\mu\pm\sigma\rightarrow\pm1$\tnote{***}  &$\mu\pm\sigma\rightarrow\pm1$
\\
\hline
\end{tabularx}
\label{tab:m2}
\begin{tablenotes}
    \footnotesize
    \item[*] When training of the $\iNN$ model, four of the eight inputs, i.e., $\alpha$, $\Msz$, $\sigdwz$ and $\ldwz$, are prescribed with constant values. 
    \item[**] The $\iNN$ is curated after each epoch, and the one with minimum test loss is selected as the final model.
    \item[***] Here $\mu$ stands for the mean and $\sigma$ stands for standard deviation.
\end{tablenotes}
\end{threeparttable}
\end{table}
    
\subsection*{Data-driven sensitivity analyses}
In this work, Spearman's correlation was employed to quantify the correlations between the POIs ($\Hc$ and $\Pp$) and their influencing factors. Compared to other correlation metrics such as Pearson's correlation, Spearman's correlation evaluates how well the relationship between two variables can be described by a monotonic function~\cite{pearson_spearman}. The Spearman rank coefficient, denoted as $\rho$, is defined as
\begin{equation}
    \rho = 1 - \frac{6\sum d_i^{2}}{n(n^{2}-1)},
\end{equation}
where $d_i$ represents the difference in ranks of paired observations, and $n$ is the number of observations. It should be noted that Spearman's correlation coefficient is a non-linear measure, taking values in the range $[-1,1]$, and indicates both the strength and the direction of monotonic relationships. Positive values signify direct correlations, while negative values indicate inverse correlations. Furthermore, hierarchical clustering analysis was performed based on Spearman's rank-order correlation.

	The Surrogate-based permutation importance test is employed for sensitivity analysis, where the importance of an influencing factor is calculated by quantifying the increase in the deviation in the prediction of the properties of interest after shuffling input values of the factors. An important feature is then the one with a vastly increased error of prediction after the permutation. As the prerequisite, the $\fNN$ model with optimized hyperparameters was trained for coercivity $\Hc$ and inhomogeneity factor $\Pp$. 

 \subsection*{Machine learning model for inverse design}

 For inverse design, another NN model ($\iNN$) is employed to map the queried properties ($\Yv=\{\Hc, \Pp\}$) directly onto the design parameters (here $\Xv=\{L,d,\ws,\wz\}$ since micromagnetic descriptors are prescribed). The main challenge is to define a loss function that accurately identifies the correctness of a sole inversely-predicted $\Xv$ while avoiding interference from multiple correct ones, as various nanostructures can result in a similar $\Hc$ and $\Pp$. To overcome this challenge, a modified loss function model is employed as \cite{kumar2020inverse}
 \begin{equation}
     \mathcal{L} = \min_{\nvv}\frac{1}{n}\sum^n
     \left(\lambda\underbrace{\|\imod_{\nvv}[\Yv]-\Xv\|^2}_{\text{SE}_i}+\underbrace{\|\fmod_{\wv}[\imod_{\nvv}[\Yv]]-\Yv\|^2}_{\text{SE}_f}\right)\quad\text{with}\quad\lambda\geq0,\label{eq:lossi}
 \end{equation}
 where $\imod$ stands for the $\iNN$ model with a set of model parameters $\nvv$, 
$\text{SE}_i$ and $\text{SE}_f$ are the squared errors of inverse prediction and forward verification, respectively. In this work, the forward verification ($\fmod[\imod[\Yv]]$) is implemented by the $\fNN$ model (details described in the section \textit{Machine learning (ML) model for forward prediction}) rather than micromagnetic simulations. This strategy, known as a gradient-based method, is also used in the other inverse design works \cite{kumar2020inverse,peng2024data}. 
The reason is not just for computational cost-efficiency, but also for performing automatic-differentiation-based back-propagation on loss function (i.e., $ \frac{\partial\mathcal{L}}{\partial \Xv}=\frac{\mathcal{L}}{\partial \Yv}\frac{\partial \Yv}{\partial \Xv}$) in the Adam optimizer in the training framework \cite{kingma2014adam}. $\lambda$ is a regularization parameter for tuning the training performance, in this work $\lambda=0.02$. It is worth noting that the $\fNN$ model remains the same as it was used for forward data-driven analyses, but has $\alpha$, $\Msz$, $\sigdwz$, and $\ldwz$ prescribed to fetch the $\iNN$ outputs. The optimized hyperparameters are listed in \tabref{tab:m2}. 

It is worth noting that a biased ReLU activation was applied at the $\iNN$ output as a consistency filter to enforce the non-negativity of the inversely designed quantities, i.e., nanostructural descriptors $L,d,\ws$ and $\wz$, as the schematic presented in \subfigref{fig5}{a}. This activation introduces a set of learnable bias parameters that effectively shift the rectification threshold, ensuring that the final outputs remain strictly non-negative while allowing the model to adapt the lower bound during training. A similar method has also been employed in other works to ensure non-negative outputs \cite{wang2025real, asheri2023data}.

	\section*{Data Availability}
	The authors declare that the data supporting the findings of this study are available within the paper. The simulation results, supplementary data, and utilities are curated in the online dataset (DOI: \url{xx.xxxx/zenodo.xxxxxxx}).
	
	\section*{Acknowledgements}
	The authors acknowledge the financial support of the German Science Foundation (DFG) in the framework of the Collaborative Research Centre Transregio 270 (CRC-TRR 270, project number 405553726, sub-projects A05, A06, A10, B13, Z01, Z-INF) and 361 (CRC-TRR 361, project number 492661287, sub-projects A05), and the Research Training Groups 2561 (GRK 2561, project number 413956820, sub-project A4). The authors also greatly appreciate the access to the Lichtenberg II high-performance computer by the NHR Center NHR4CES@TUDa (funded by the German Federal Ministry of Education and Research and the Hessian Ministry of Science and Research, Art and Culture), High-performance computer HoreKa by the NHR Center NHR@KIT (funded by the German Federal Ministry of Education and Research and the Ministry of Science, Research and the Arts of Baden-Württemberg, partly funded by the DFG), and the GPU Cluster from the CRC-TRR 270 sub-project Z-INF (funded by the DFG). The NHR4CES Resource Allocation Board allocates computing time on the HPC under the project ``special00007''. Y. Yang and M. Fathidoost highly thank Dr. Binbin Lin and Dr. Xiang-Long Peng for providing the strategy for training the forward and inverse ML models. Y. Yang and D. Ohmer also highly thank Prof. Michael Farle and Prof. Lambert Alff for their intensive consultation and discussion.
	
\section*{Competing Interests}
	The authors declare no competing financial or non-financial interests.

\section*{Author Contributions}\label{Method}
Conceptualization: B.-X.X. and Y.Y.; methodology: Y.Y., B.-X.X. and M.F.; software: Y.Y., and M.F.; characterization/validation: E.A. and L.M-L.; investigation: Y.Y., P.K. and E.F.; formal analysis: Y.Y., P.K. and M.F.; resources, K.S., O.G. R.X. and H.Z.; data curation, Y.Y.; writing—original draft preparation: Y.Y., P.K. and M.F.; writing—review and editing: all authors; visualization: Y.Y., M.F.; supervision, B.-X.X. and H.Z.; consultation and discussion: K.S. and O.G.; funding acquisition, K.S., H.Z., O.G., L M-L. and B.-X.X. All authors have read and agreed to the published version of the manuscript.

 \clearpage
\bibliographystyle{naturemag}
\bibliography{reference, ref_data, ref_smco}
\clearpage

\end{document}